\documentclass[12pt]{article}
\usepackage{ssi}
\usepackage{times}

\usepackage{subfigure}

\usepackage{graphicx}
\usepackage{wrapfig}

\def\ET{\mbox{$\mathrm{E}_\mathrm{T}$}}
\def\pT{\mbox{$p_T$}}
\def\sqrtsNN{\mbox{$\sqrt{s_\mathrm{NN}}$}}
\def\sqrts{\mbox{$\sqrt{s}$}}
\def\Npart{\mbox{$\mathrm{N}_\mathrm{part}$}}
\def\Nbinary{\mbox{$\mathrm{N}_\mathrm{binary}$}}

\def\Temp{\mbox{$\mathrm{T}$}}
\def\TC{\mbox{$\mathrm{T}_\mathrm{C}$}}
\def\epsC{\mbox{$\epsilon_\mathrm{C}$}}

\def\muB{\mbox{$\mu_\mathrm{B}$}}

\def\LambdaQCD{\mbox{$\Lambda_\mathrm{QCD}$}}
\def\epsBj{\mbox{$\epsilon_\mathrm{Bj}$}}

\def\Tchem{\mbox{$\mathrm{T}_\mathrm{chem}$}}

\def\GeVfmCubed{\mbox{$\mathrm{GeV}/\mathrm{fm}^3$}}

\def\LuminosityUnits{\mbox{$\mathrm{cm}^{-2}\mathrm{sec}^{-1}$}}

\def\pizero{\mbox{$\pi^0$}}

\def\pbar{\mbox{$\bar\mathrm{p}$}}

\def\pbarp{\mbox{$\pbar+\mathrm{p}$}}

\def\jpsi{\mbox{$\mathrm{J/}\psi$}}

\def\dNchdeta{\mbox{$dN_{ch}/d\eta$}}

\def\vtwo{\mbox{$v_2(\pT)$}}
\def\dedx{\mbox{$\mathrm{dE/d}x$}}
\def\rhoglue{\mbox{$\rho_\mathrm{glue}$}}

\def\RAA{\mbox{$R_{AA}(\pT)$}}
\def\Qsubs{\mbox{$Q_s^2$}}

\begin{document}

\title{Measurements of High Density Matter at RHIC}

\author{Peter Jacobs\thanks{pmjacobs@lbl.gov}
\vskip 0.5in 
\noindent
Lawrence Berkeley National Laboratory, \\
Berkeley, CA 94720 \\[0.4cm]
}

\maketitle
\begin{abstract}%
\baselineskip 16pt 

QCD predicts a phase transition between hadronic matter and a Quark
Gluon Plasma at high energy density. The Relativistic Heavy Ion
Collider (RHIC) at Brookhaven National Laboratory is a new facility
dedicated to the experimental study of matter under extreme
conditions. I will discuss the first round of experimental results
from colliding heavy nuclei at RHIC and our current understanding of
the state of matter generated in such collisions, concentrating on
partonic energy loss and jet quenching as a probe of the medium.

\end{abstract}


\section{Introduction}

At high temperature or baryon density, hadronic matter dissolves into
a soup of its constituent quarks and gluons. For an asymptotically
free field theory such as QCD, the state of matter at high energy
density is simple\cite{CollinsPerry}: long range (low
momentum) interactions are screened, and short range (high momentum)
interactions are weak, leading to an ideal gas equation of state in
the high energy density limit. At temperature $\Temp\gg\LambdaQCD$ matter is
a gas of deconfined, weakly interacting quarks and gluons (the
``Quark-Gluon Plasma'', or QGP), whereas at $\Temp\ll\LambdaQCD$
quarks and gluons are confined and matter consists of strongly
interacting hadrons.

\begin{wrapfigure}{l}{0.5\textwidth}
\includegraphics[width=0.48\textwidth]{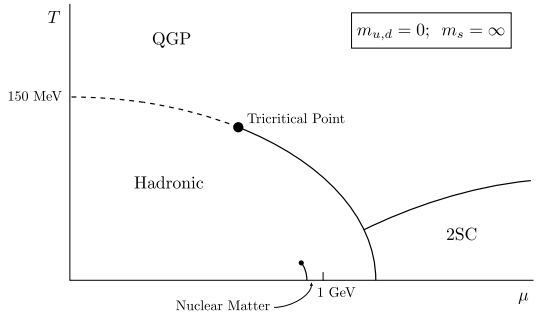}
\caption{Example of the QCD phase diagram (temperature T vs. baryochemical potential \muB) for two 
massless flavors \cite{WilczekRajugopal}. Chiral symmetry is broken in
the hadronic phase. ``2SC'' is a color-superconducting phase.}
\label{figPhaseDiagram}
\end{wrapfigure}

The QCD phase diagram has a complex structure
(Fig. \ref{figPhaseDiagram}). At low temperature and baryon
density the phase is hadronic (confined phase) and chiral symmetry is
broken. Color-superconducting and other phases may exist at high
baryon density and low temperature
\cite{WilczekRajugopal}, whereas at high temperature the quarks and
gluons are deconfined and chiral symmetry is restored. The early
universe descended from high T at extremely small \muB. Neutron star
cores have high \muB\ and very low T.

\begin{wrapfigure}{l}{0.5\textwidth}
\includegraphics[width=0.48\textwidth]{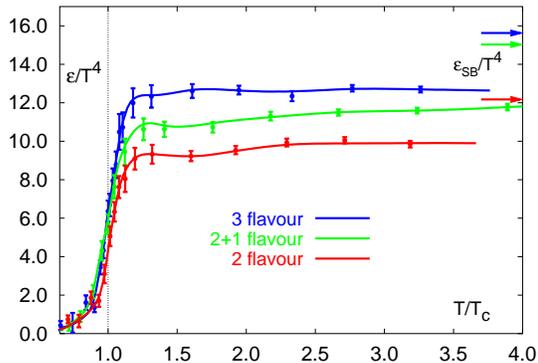}
\caption{Lattice calculation of energy density vs temperature for 
two- and three-flavor QCD \cite{Karsch}.}
\label{figLattice}
\end{wrapfigure}

First-principles calculations of finite temperature QCD can only be
carried out numerically on the lattice
\cite{KanayaLattice}. Fig. \ref{figLattice} shows a recent lattice
calculation of the energy density $\epsilon$ as function of
temperature for two- and three-flavor QCD. $\epsilon$ exhibits a sharp
rise in the vicinity of the critical temperature $T_C$, indicating a
rapid change in the density of underlying degrees of freedom. The
ideal gas Stefan-Boltzmann limit $\epsilon_\mathrm{SB}$ has not yet
been achieved at $T\sim4T_C$. Putting in physical values, $\TC\sim175$
MeV, resulting in critical energy density $\epsC=(6\pm2)\TC^4\sim1$
\GeVfmCubed. This value should be kept in mind for
comparison to conditions achieved in laboratory experiments.

The order of the deconfinement phase transition can be determined in
some limiting cases\cite{KanayaLattice}. It is first order for pure
gauge and for three light quarks, second order for two light and one
heavy quark. For physical quark masses the order of the transition, or
indeed whether it is a smooth cross over, has not been determined. The
extension of lattice calculations to $\muB\neq0$ is a long-standing
problem, but there has been significant recent progress in determining
the phase boundary and equation of state for finite
\muB\cite{FodorQM02}.
 
In the early universe, the confinement transition
(QGP$\rightarrow$hadrons) at very low \muB\ occurred about 10
microseconds after the big bang\cite{Boyanovsky}. A strongly first
order phase transition may have generated primordial black holes,
strange quark nuggets, or local baryon asymmetries affecting
primordial nucleosynthesis, though no relics of this transition have
been observed thus far. In the cores of neutron stars, a QGP phase at low
temperature and high \muB\ may generate observable millisecond pulsar
phenomena\cite{GlendenningWeber}.

It is natural to ask whether the deconfinement and chiral symmetry
restoration transitions can be studied in accelerator-based
experiments. We require a system having temperature of order the pion
mass in equilibrium over a large volume. The best possibility to
accomplish this is through the collision of heavy nuclei at the
highest possible energy.

However, nuclear collisions at high energy are highly dynamic. Even if
a hot, equilibrated system is created early in the collision, it will
immediately expand and cool. If a QGP is formed it will
hadronize after a brief period, and signatures from the deconfined
phase may be masked by those from the hot, interacting hadron gas. The
central issue is to find and measure those experimental observables that
are sensitive to the state of matter early in the
collision\cite{MullerHarris}.

Broadly speaking, several epochs in the evolution of high energy
nuclear collisions can be sketched. Immediately following the
collision is a brief formation time $\tau<1$ fm/c, during which large
momentum transfer (short-distance) processes occur. It is also during
this period that the highest energy density is achieved and the QGP
may result. After an expansion time of perhaps a few fm/c the plasma
hadronizes into a dense, interacting hadron gas. Further expansion and
cooling causes the temperature to fall to the point at which inelastic
collisions among hadrons are no longer common ({\it chemical
freezeout}) and the relative populations of the various long-lived
hadron species are established. Elastic collisions occur until {\it
kinetic freezeout}, after which point the hadrons fly effectively
undisturbed to the detectors. Thus, the relative populations of stable
hadrons reflect the conditions at chemical freezeout, whereas their
momentum spectra reflect the conditions at kinetic freezeout. More
penetrating probes (dileptons, direct photons, jets, heavy
quarks,\dots) may carry information on the conditions around the time
of their formation earlier in the collision.

Collisions of heavy nuclei at ultra-relativistic energies have been
studied in fixed target experiments at the Brookhaven AGS and CERN SPS
for the past 15 years. In 1999, the Relativistic Heavy Ion Collider
(RHIC) at Brookhaven National Laboratory was commissioned, bringing
into operation the first facility largely dedicated to the study of
matter at high energy density. 

I will first describe the RHIC machine and experiments and then
discuss some of the main results from its heavy ion physics program. I
will briefly touch on ``soft physics'' observables ($\pT<\sim2$ GeV/c) and
summarize what has been learned from them. The bulk of the review
concentrates on the major new development at RHIC: jet production and
indications of partonic interactions with dense matter, which may
directly probe the energy density achieved early in the collision.


\section{The Relativistic Heavy Ion Collider}

RHIC consists of two independent superconducting rings, 3.8 km in
length. It has enormous flexibility in beam masses and energies, with
capability to collide gold ions from \sqrtsNN=20 to 200 GeV per
nucleon pair, protons up to \sqrts=500 GeV, and asymmetric systems,
most importantly protons and deuterons with heavy nuclei. The top CM
energy for heavy nuclei is a factor 10 larger than for the fixed
target experiments, extending significantly the statistical and
transverse momentum reach of many observables and opening up new
channels. RHIC is also the first polarized proton collider, opening
new opportunities to study the spin content of the proton, in
particular the contribution of the gluon at low $x_\mathrm{Bj}$. More
details of the machine and its first year performance can be found in
Ref. (\citenum{Roser}).

Design luminosity for Au+Au at 200 GeV is ${\cal L}=2\cdot10^{26}\
\LuminosityUnits$, giving an interaction rate of about 2 kHz. While this 
luminosity appears to be tiny relative to other modern colliders,
recall that the rate for hard processes such as jet production in
nuclear collisions scales as $\sim{A}^2$ ($A$=atomic mass), so that
hard process rates in Au+Au at design luminosity are the same as at a
proton collider with ${\cal L}\approx10^{31}\ \LuminosityUnits$. The
design luminosity for p+p collisions at 500 GeV is $2\cdot10^{32}\
\LuminosityUnits$, with an interaction rate of about 8 MHz. 
Design polarization for p+p is 70\%.

RHIC has six intersection regions, of which four are currently
instrumented with experiments\cite{RHICExperiments}.  PHENIX consists
of an axial field magnet and four independent spectrometers: two at
midrapidity containing tracking, ring imaging Cerenkov counters, time
of flight, and electromagnetic calorimetry, which are optimized for
precision lepton, photon and hadron measurements, and two
forward muon arms. STAR has conventional collider detector geometry,
with a large solenoidal magnet, Time Projection Chamber for tracking,
large coverage EM calorimetry, and an inner silicon-based
tracker. STAR is designed for hadron, jet, lepton and photon
measurements over large acceptance, as well as studies of event-wise
fluctuations in high multiplicity nuclear collisions. BRAHMS consists
of two small acceptance spectrometers for inclusive identified hadron
measurements over wide phase space. PHOBOS has very wide phase space
coverage for charged particles using silicon detectors, and a
mid-rapidity spectrometer based on a dipole magnet, silicon tracking
and Time of Flight. STAR and PHENIX each have about 450 collaborators,
whereas PHOBOS and BRAHMS each have fewer than 100.

RHIC had a brief commissioning run in 1999. The data reported
here are from a Au+Au run at \sqrtsNN=130 GeV in 2000 and Au+Au and
polarized p+p ($\sim$15\% polarization) runs at
\sqrtsNN=200 GeV in 2001-2. The integrated luminosity is
about 80 $\mu\mathrm{b}^{-1}$ for Au+Au and 1 pb$^{-1}$ for p+p. A
comprehensive view of the physics program of RHIC and what has been
achieved thus far can be found in the proceedings of the recent
Quark Matter conferences\cite{QM01,QM02}.

\begin{wrapfigure}{l}{0.5\textwidth}
\vspace{-3mm}
\includegraphics[width=0.48\textwidth]{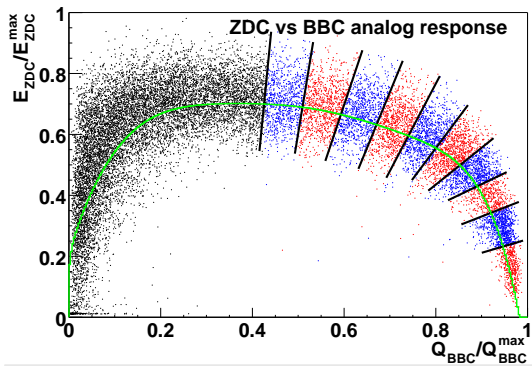}
\vspace{-3mm}
\caption{Event characterization\cite{PHENIXmult}: correlation of ZDC signal (vertical) with 
multiplicity.}
\label{figTrigger}
\vspace{-6mm}
\end{wrapfigure}

Nuclei are extended objects, and high energy nuclear collisions can be
characterized experimentally as head-on (``central'') or glancing
(``peripheral''). For this purpose all experiments have identical
forward hadron calorimeters (Zero Degree Calorimeters, or
ZDCs\cite{ZDC}), situated $\pm$18 m downstream of the interaction
region beyond the first machine dipole magnets. The ZDCs measure the
forward-going ``spectator'' neutrons which did not scatter
significantly. The correlation of the ZDC signal with the event
multiplicity serves to characterize the geometry of the
event. Peripheral collisions have few spectator neutrons and low
multiplicity. More central collisions generate more spectator neutrons
and higher multiplicity, while the most central, head-on collisions
again generate few spectator neutrons and the highest
multiplicity. Fig. \ref{figTrigger} shows this correlation from the
PHENIX experiment, the other experiments use similar
distributions. Based on this correlation the total geometric cross
section can be divided into bins containing a certain percentile of
``centrality'', as shown by the lines on the figure.

Related quantities which also characterize the centrality of the
event are the number of incoming participating nucleons (\Npart), and
the number of equivalent binary nucleon-nucleon collisions
(\Nbinary). \Npart\ and \Nbinary\ are calculated for each centrality
bin using the Glauber formalism and a realistic model of the nuclear
geometry\cite{Kharzeev}. Phenomenologically it has been found that
total particle production scales roughly as \Npart, whereas the rate
of hard processes will scale as \Nbinary\ in the absence of nuclear
effects. Such scaling rules can be used to uncover the new
physics present in nuclear collisions. For instance, the violation of
\Nbinary\ scaling at high \pT\ indicates significant effects of the nuclear medium on high \pT\
processes. I will return to this point below.

\section{Soft Physics}

\begin{wrapfigure}{l}{0.38\textwidth}
\vspace{-3mm}
\includegraphics[width=0.35\textwidth]{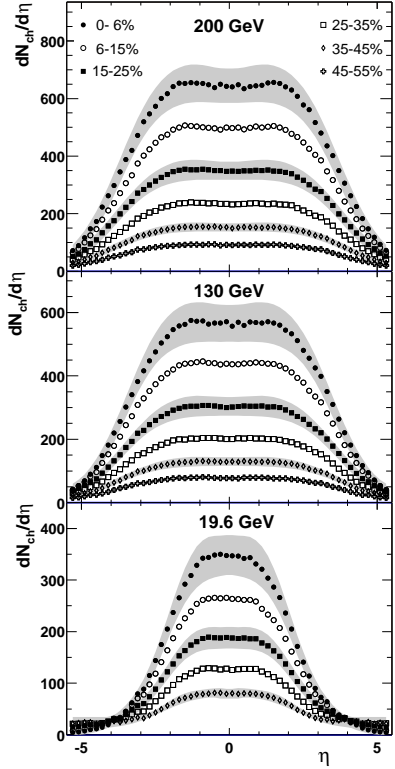}
\vspace{-3mm}
\caption{\dNchdeta\ for various collision centralities 
over the full RHIC phase space\cite{PhobosdNchdeta}.}
\label{dNchdeta}
\vspace{-2mm}
\end{wrapfigure}

Fig. \ref{dNchdeta} from the PHOBOS
Collaboration\cite{PhobosdNchdeta} shows charged particle multiplicity
distributions \dNchdeta\ vs $\eta$\cite{eta} over the full RHIC phase
space, for Au+Au collisions at all three collision energies studied so
far. The longitudinal phase space growth with increasing energy is
apparent, as well as the increase in multiplicity for more central
collisions. For the most central collisions at \sqrtsNN=130 GeV, 4200
charged particles are produced in the full phase space.

Figure \ref{dNchdeta} exhibits a central plateau near $\eta\sim0$,
indicating that the system has approximate longitudinal boost
invariance. More detailed considerations of identified particle
spectra show that boost invariance holds over a rather smaller region
$\Delta{y}\sim1$\cite{UllrichQM02}. Bjorken studied boost invariant
hydrodynamics\cite{BjorkenHydro} and derived a useful pocket formula
for the energy density achieved in the central region:

\begin{eqnarray}
\epsBj=\frac{1}{\pi{R_A}^2\tau}\frac{d\ET}{dy}\approx
\frac{1}{\pi{R_A}^2\tau}\langle\pT\rangle\frac{3}{2}\frac{dN_{ch}}{d\eta},
\label{eqBjEpsilon}
\end{eqnarray}

\noindent
where \ET\ and \dNchdeta\ are the transverse energy and
multiplicity density (Fig. \ref{dNchdeta}). $R_A$ is the nuclear radius and
$\tau$ is the formation time, typically taken as $\sim1$ fm/c, after
which the hydrodynamic description is valid. PHENIX measured \ET\ for
central Au+Au collisions at \sqrtsNN=130 GeV and derived an energy
density of 4.6
\GeVfmCubed\cite{PhenixET}, well above 1
\GeVfmCubed\ required for the deconfinement transition as calculated 
on the lattice. Conditions appear to be favorable to form a QGP in
nuclear collisions at RHIC, though under the assumption that an
equilibrated system is established after a time $\sim1$ fm/c.

Scaling of global observables such as multiplicity with collision
centrality and \sqrts\ has been discussed extensively in the context
of gluon saturation (e.g. Ref. \citenum{Kharzeev} and references
therein). A projectile parton at sufficiently low $x$ interacts
coherently with all partons in the target that lie within a transverse
area $1/Q^2$, with cross section $\sigma\sim\alpha_s/Q^2$. If the
target is a heavy nucleus, its finite thickness will increase the
parton density seen by the probe: nuclei are ``blacker'' than
protons. The $Q^2$ scale at which the nucleus becomes dense to the
probe is given by the saturation condition\cite{Kharzeev}:

\begin{equation}
\Qsubs\sim\alpha_s(\Qsubs)\frac{xG_A(x,\Qsubs)}{\pi{R}_A^2}\sim{A}^\frac{1}{3},
\label{eqQs}
\end{equation}

\noindent
where $xG_A\sim{A}$ is the nuclear gluon distribution and
$R_A\sim{A}^\frac{1}{3}$ is the nuclear radius. For central Au+Au
collisions at RHIC $\Qsubs\sim2$ GeV$^2$, so that $\alpha_s(\Qsubs)$
is small: in the saturation regime the coupling is weak while the
gluon density is high, leading to classical, nonlinear dynamics. The
growth of \Qsubs\ with target thickness leads e.g. to definite
predictions for the dependence of multiplicity on collision
centrality\cite{Kharzeev}.

\begin{figure}
\centering
\mbox{
\subfigure[Multiplicity($\eta\sim0$) per participant pair vs. centrality\cite{PhobosSaturation}.]{
\includegraphics[width=0.48\textwidth]{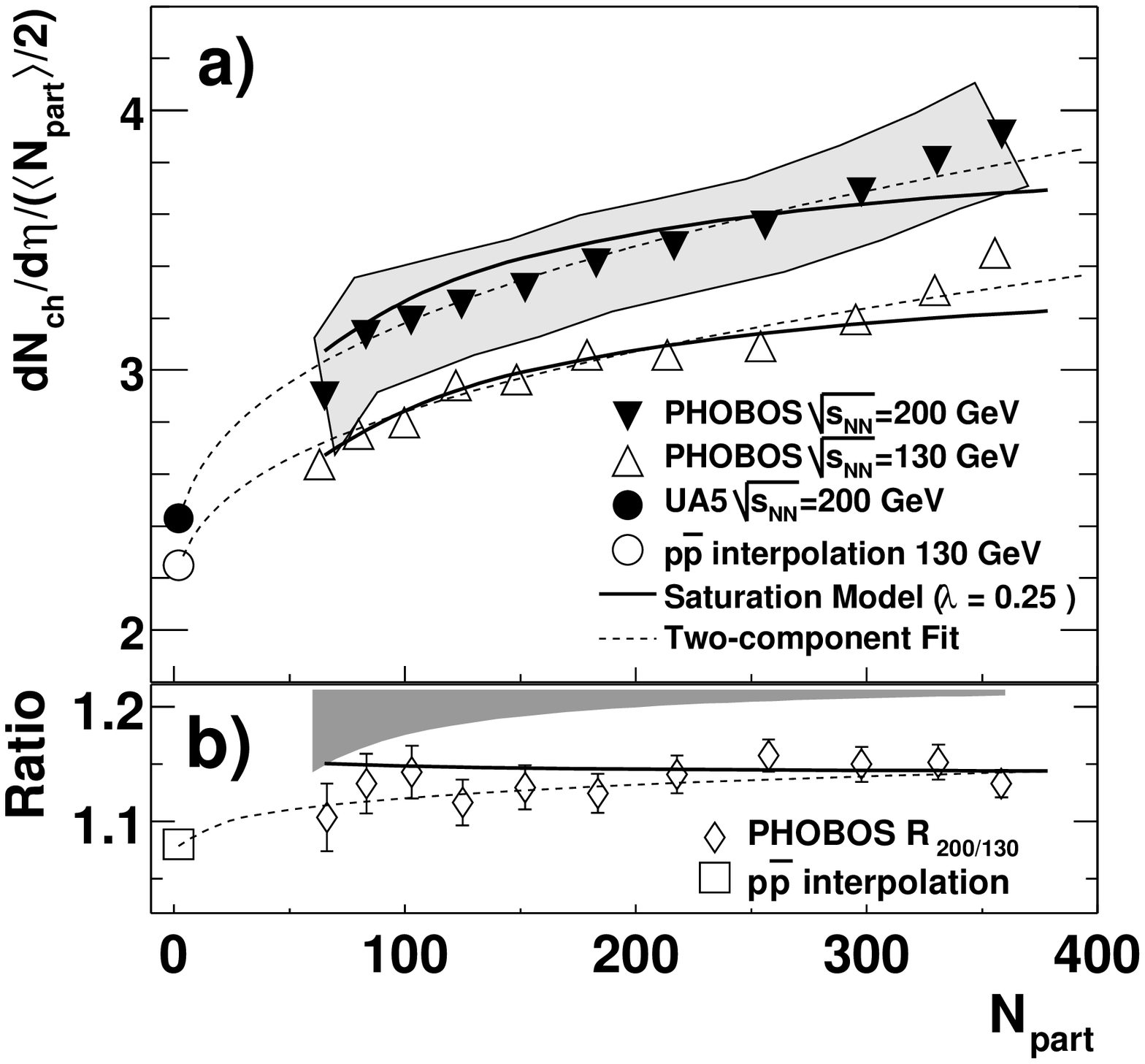}
\label{PhobosMultNpart}}
\subfigure[Universal scaling of total multiplicity vs. \sqrts\ for $e^+e^-$, p+p and 
Au+Au\cite{SteinbergQM02}.]{
\includegraphics[width=0.48\textwidth]{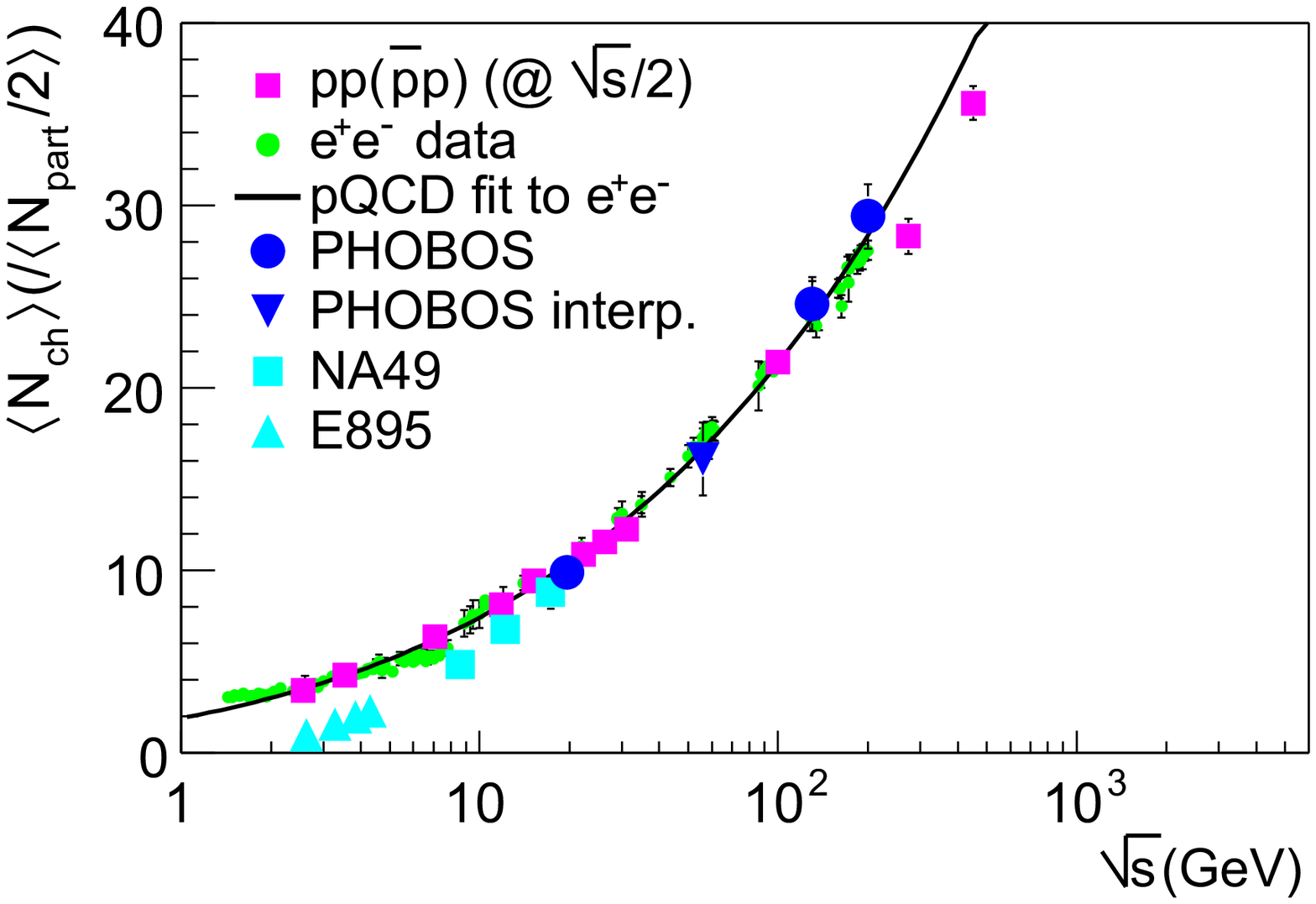}
\label{PhobosMultSqrts}}
}
\caption{}
\label{PhobosMult}
\end{figure}

Fig. \ref{PhobosMultNpart} from PHOBOS
illustrates this dependence for Au+Au collisions at 130 and 200
GeV (central collisions correspond to large number of participants
\Npart, peripheral to small \Npart). The data are compared to
both the saturation model and a more conventional ``two component''
model, which parameterizes the centrality dependence of multiplicity
in terms of soft production (scaling with \Npart) and minijet
production (scaling with \Nbinary). The data are consistent with both
models and cannot resolve the small difference between them. While
this is a very active area of research, no clear evidence of
saturation effects in nuclear collisions has been found yet at RHIC.

Fig. \ref{PhobosMultSqrts} from PHOBOS shows a universal scaling of
the total multiplicity with
\sqrts\ for $e^+e^-$, p+p and Au+Au\cite{SteinbergQM02}. (The p+p 
data are corrected for energy carried by the leading particles.)  This
is at first sight surprising, given the very different nature of the
various systems, but may point to universal features of particle
production in high energy collisions (though it should also be noted
that the evolution with \sqrts\ of the mean transverse momentum shows
no such scaling among the various systems\cite{UllrichQM02}).

\begin{figure}
\includegraphics[width=0.95\textwidth]{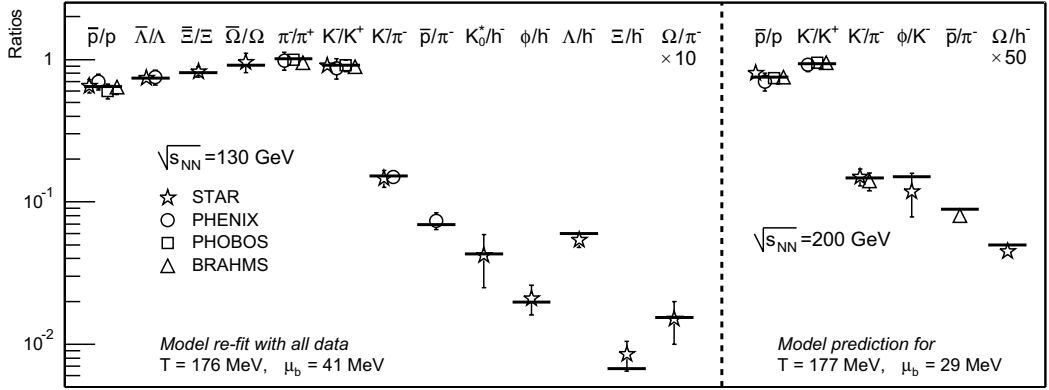}
\caption{Population ratios of stable hadrons and statistical 
model fit\cite{PBMChemical,UllrichQM02}.}
\label{ChemicalFit}
\end{figure}

As argued above, the relative population of the stable hadrons
is fixed at chemical freezeout. Equilibrium at that point should be
evident in the measured population ratios. Fig. \ref{ChemicalFit} shows
population ratios measured at mid-rapidity for a wide variety of
non-strange and strange hadrons, with values varying over three orders
of magnitude. Also shown are results from a statistical model fit to
the data\cite{PBMChemical}. The model consists of a partition function
containing contributions from all baryons and mesons with mass less
than 2 GeV, with two free parameters (chemical freezeout
temperature \Tchem\ and baryochemical potential
\muB) and constraints due to strangeness and charge conservation. The
resulting fit from this and similar models is excellent, with
extracted parameters $\Tchem\sim170$ MeV and $\muB\sim25-50$ MeV, close
to the critical temperature for the deconfinement phase transition
calculated on the lattice. A hadronic gas cannot exist at a higher
temperature, suggesting that equilibrium is in fact established {\it
prior} to hadronization, in the deconfined phase.

I cannot hope to address all the soft physics observables that have
been studied at RHIC. To conclude this section I will summarize a few
of the main results, more detailed discussions can be found in Refs.
(\citenum{QM02},\citenum{UllrichQM02}).

\begin{itemize}

\item Low baryon density: antibaryon/baryon ratios at midrapidity 
are $\sim$0.6-1.0. The system is close to baryon free, similar to the
early universe, but not precisely so. Finite baryon number is
transported $\Delta{y}\sim5.5$ rapidity units from the beam.

\item There are strong indications that hadronic chemical equilibrium has been achieved 
at a temperature near the lattice critical temperature.

\item Hydrodynamic calculations describe well the transverse momentum spectra and ``elliptic flow'' 
(see below). The mass dependence of these observables, which is a
sensitive test of the hydrodynamic picture, is described in detail.

\item The energy density achieved early in the collision is estimated to be $\sim5$ \GeVfmCubed.

\item Identical two-particle correlations (Hanbury Brown-Twiss correlations) are sensitive to 
the space-time evolution of the source. The extracted source radii and
duration of freezeout show no significant increase relative to lower
energy nuclear collisions. Such correlations measure only a piece of a
dynamically expanding source and these results may indicate a very
explosive expansion resulting from high early pressure.

\end{itemize}

\section{High \pT\ Hadrons and Jets}

The increase in \sqrtsNN\ for nuclear collisions at RHIC relative to
fixed target experiments opens up new channels to probe the dense
medium generated in the collision. Jets with $\ET\sim40$ GeV and
higher are produced in sufficient numbers to provide robust
observables. The measurement of jets in nuclear collisions poses a
special problem, however: while the presence of a hard scattering in a
nuclear collision can be detected (though in a biased way) via high
\pT\ leading hadrons, the huge soft multiplicities seen in
Fig. \ref{dNchdeta} contaminate any finite jet cone, spoiling the jet
energy measurement. At sufficiently high \ET\ this effect may be minor
(e.g. nuclear collisions at the LHC), but for the jet \ET\ currently
accessible at RHIC it is fatal. We therefore restrict our
considerations to leading particles and their correlations. We show
below that hadrons with $\pT>\sim4$ GeV/c are produced dominantly
from jet fragmentation, even in the most central Au+Au collisions.

Twenty years ago Bjorken\cite{BjFermilabNote} proposed that hard
scattered partons in nuclear collisions could provide a sensitive
probe of the surrounding medium. The energy loss \dedx\ due to elastic
scattering of the partons in a Quark Gluon Plasma depends on the
temperature as $T_\mathrm{plasma}^2$ and results in a suppression of
the observed rate of jets or their leading hadrons at fixed \pT. Later
work showed that
\dedx\ from elastic scattering is negligible but that radiative energy loss 
in dense matter could be considerable\cite{JetQuench,Baier,XNfragmentation}. The energy loss
is directly sensitive to the gluon density of the medium,
\rhoglue. While not a direct signature of deconfinement, measurement
of
\rhoglue\ that is substantially larger than in cold nuclear matter\cite{ColdMatter} 
is incompatible with the presence of a
hadronic medium, thus large energy loss serves as an indirect
signature of deconfinement.

There are currently three sets of measurements which address the
question of partonic energy loss in dense matter, which I will discuss
in turn: suppression of inclusive spectra, elliptic flow (azimuthal
asymmetry) in non-central events, and correlations of high \pT\ hadron
pairs.

\subsection{High \pT: Suppression of Inclusive Spectra}

In p+p collisions at RHIC energies, hadrons with $\pT\sim4$ GeV/c
typically carry 75\% of the energy of their parent jet, leading to the
possibility that partonic energy loss in Au+Au collisions is reflected
in the suppression of leading hadrons\cite{XNfragmentation}. Hadron
suppression is measured via the nuclear modification factor:

\begin{eqnarray}
\RAA=\frac{d\sigma_\mathrm{AA}/dyd\pT^2}{\langle\Nbinary\rangle{d}\sigma_\mathrm{pp}/dyd\pT^2}
\label{eqRAA}
\end{eqnarray}

\noindent
where $\sigma_\mathrm{pp}/dyd^2\pT$ is the inclusive cross section
measured in elementary nucleon-nucleon collisions and
$\langle\Nbinary\rangle$ accounts for the geometric scaling from the
elementary to nuclear collision. \RAA\ is normalized to be unity if
Au+Au collisions are an incoherent superposition of p+p collisions. In
addition to partonic energy loss, \RAA\ may be altered by nuclear effects
such as gluon shadowing, which will reduce \RAA, and the Cronin
effect (multiple soft scattering in the initial state), which will
increase it. These effects must be disentangled using measurements in
simpler systems (in particular p+Au).

\begin{figure}
\centering
\mbox{
\subfigure{\includegraphics[width=0.58\textwidth]{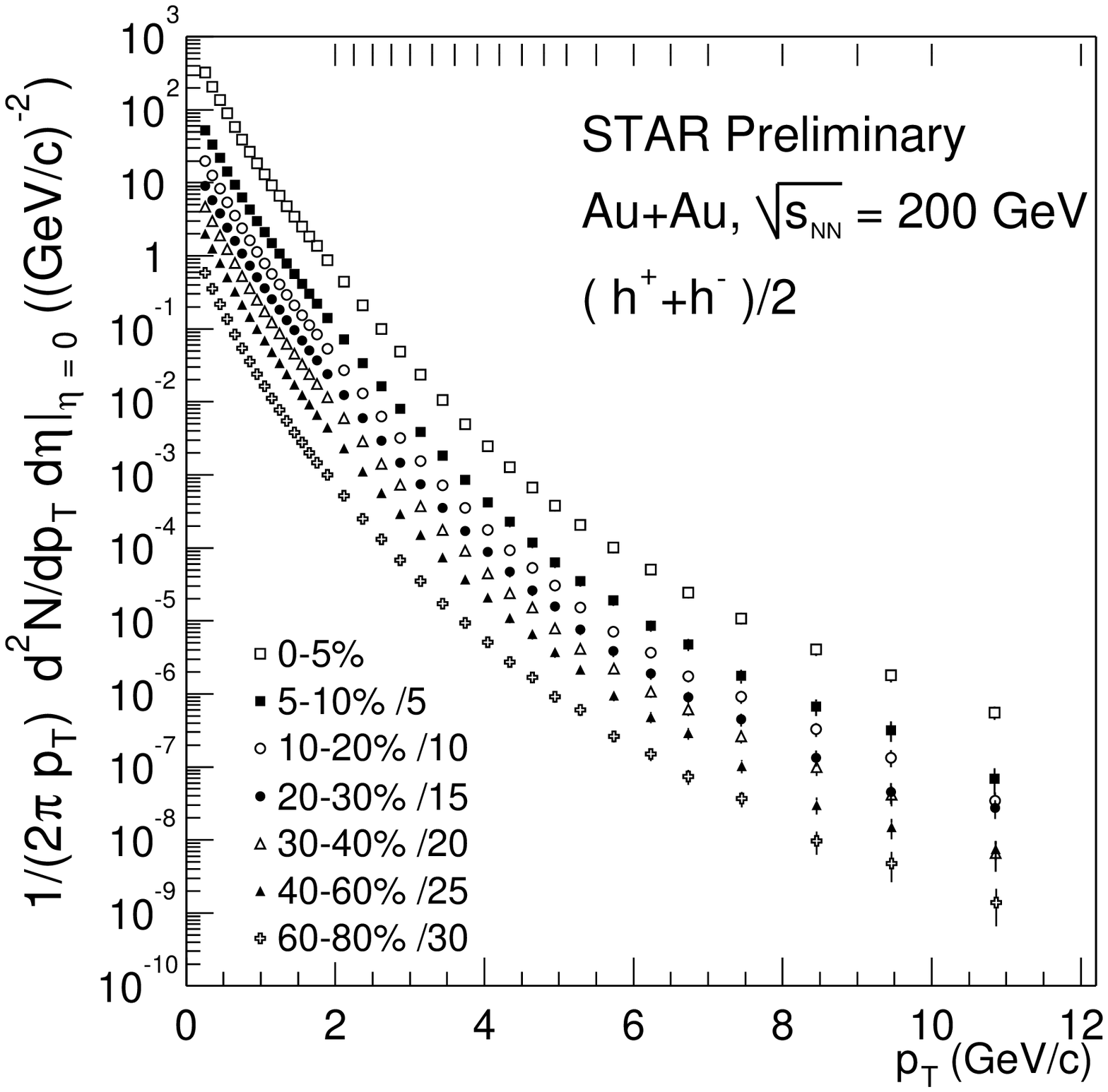}}
\subfigure{\includegraphics[width=0.38\textwidth]{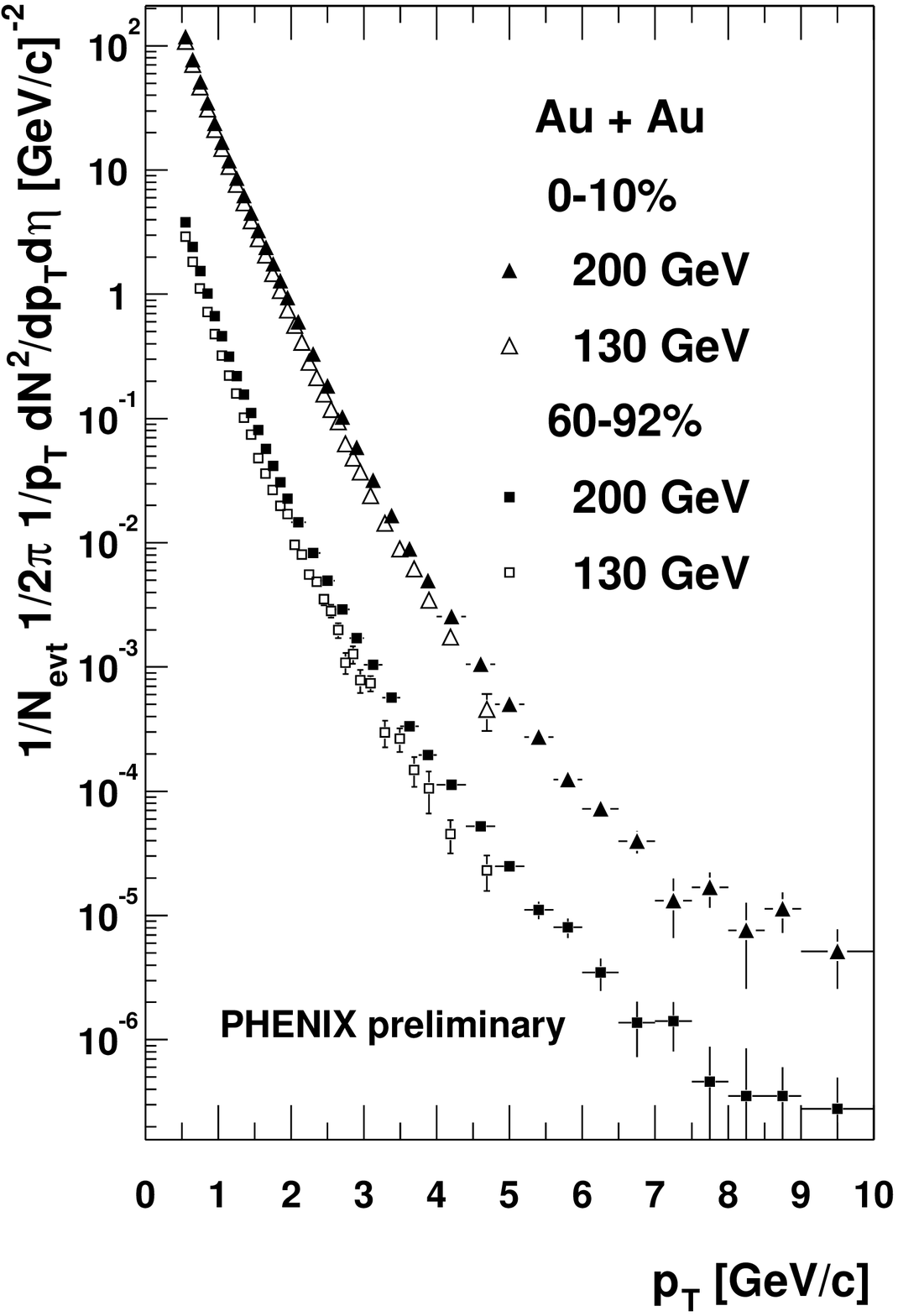}}
}
\caption{Inclusive charged hadron invariant distributions for various event centralities\cite{highpT130}.}
\label{InclusiveSpectra}
\end{figure}

\begin{figure}
\centering
\mbox{
\subfigure{\includegraphics[width=0.45\textwidth]{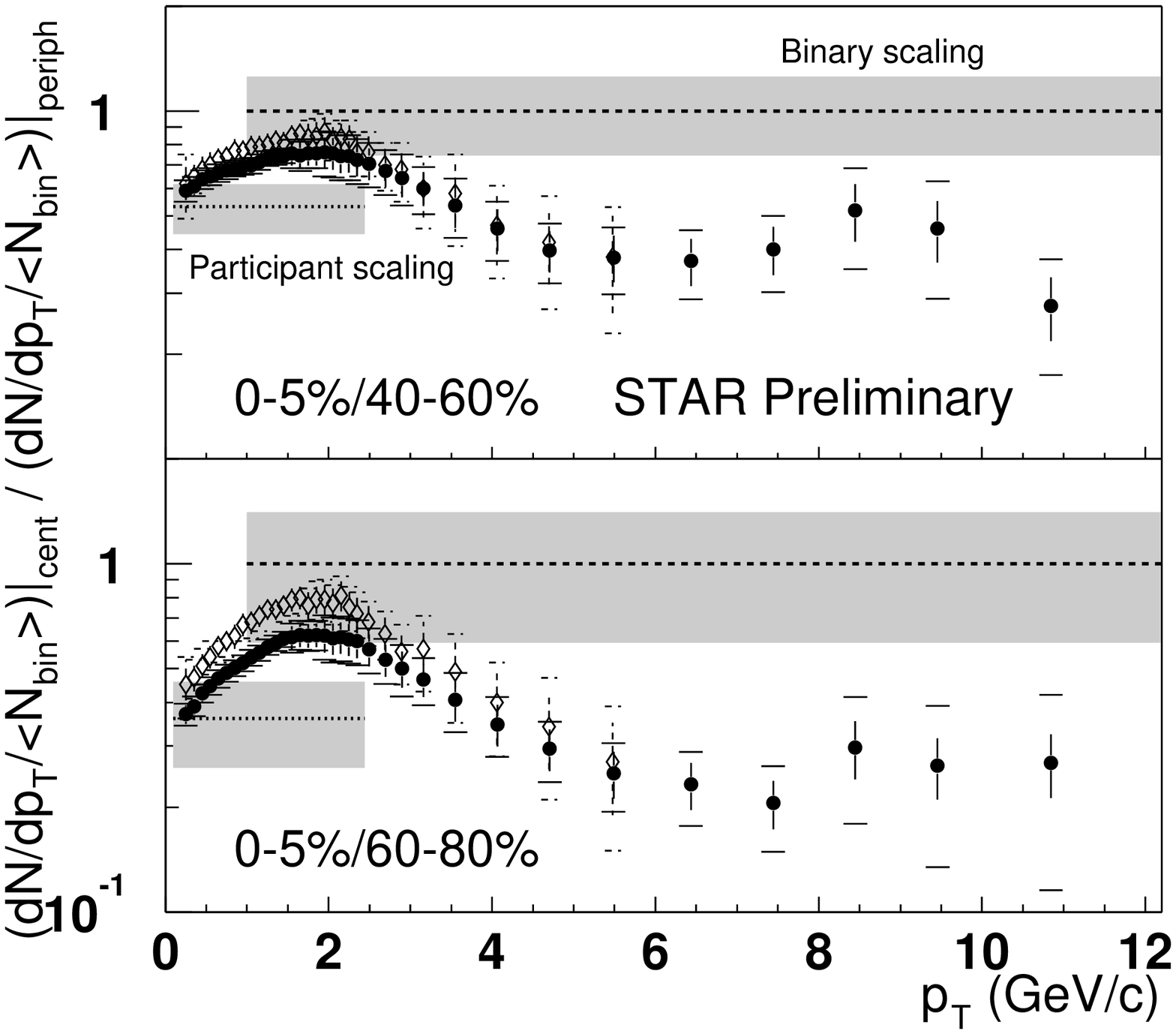}}
\subfigure{\includegraphics[width=0.45\textwidth]{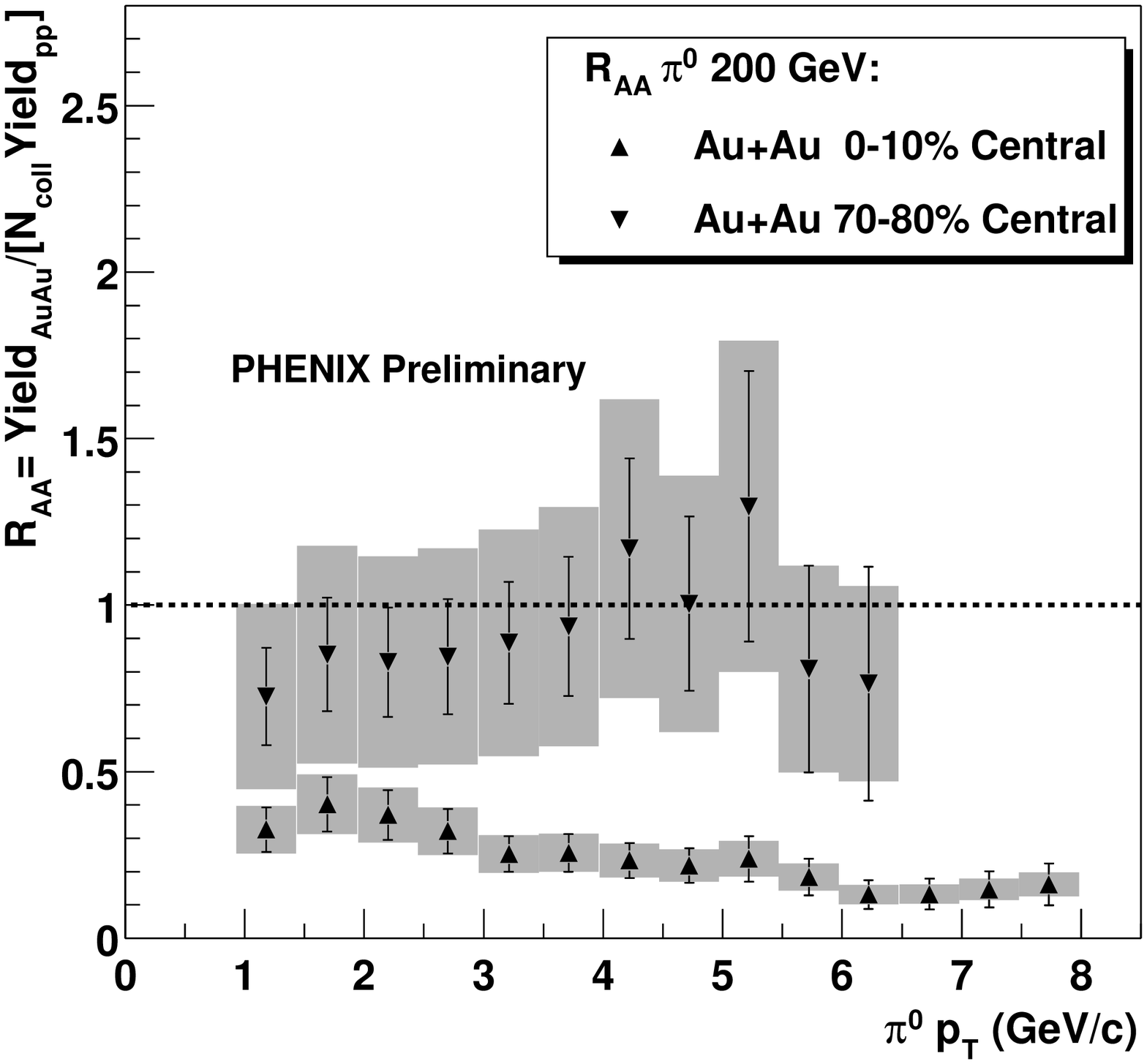}}
}
\caption{\RAA\ (eq. \ref{eqRAA} or slight modification) for charged hadrons 
(left panel, STAR\cite{JennQM02}) and \pizero (right panel, PHENIX\cite{SaskiaQM02}). }
\label{figRAA}
\end{figure}

First indication of a strong suppression of high \pT\ hadrons in
nuclear collisions was reported from the 130 GeV data by both PHENIX and
STAR\cite{highpT130}. The 200 GeV data have much higher statistics and
both collaborations have now pushed this study well into the
perturbative regime. Fig. \ref{InclusiveSpectra} shows the inclusive
charged hadron spectra as a function of centrality for Au+Au collisions at 200
GeV, extending to \pT=12 GeV. PHENIX has also shown \pizero\ spectra
from both Au+Au and p+p\cite{SaskiaQM02}. Fig. \ref{figRAA} shows
\RAA\ for these data. Suppression factors of 4-5 are observed in central 
collisions for both
\pizero{s} and charged hadrons, with weak, if any, dependence on \pT\
above $\pT\sim5$ GeV/c. At asymptotically high jet \ET\ the relative
energy loss should be negligible and \RAA\ should return to unity. No
evidence of this limiting high energy behavior is seen.

\begin{wrapfigure}{l}{0.5\textwidth}
\vspace{-3mm}
\includegraphics[width=0.6\textwidth,angle=270]{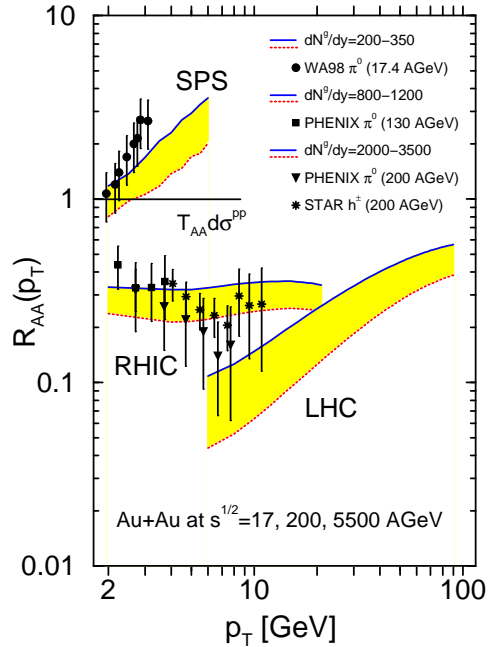}
\vspace{-3mm}
\caption{pQCD calculation of hadron suppression\cite{VitevRAA}, 
compared to data from Fig. \ref{figRAA}.}
\label{VitevRAA}
\end{wrapfigure}

Fig. \ref{VitevRAA} shows a pQCD calculation\cite{VitevRAA}
incorporating Cronin enhancement, nuclear shadowing and energy loss,
compared to the data for central collisions from Fig. \ref{figRAA} as
well as lower energy fixed target data. The RHIC data are compatible
with a gluon rapidity density $dN^g/dy\sim800-1200$, several times
that of cold nuclear matter. The calculation predicts a suppression
factor independent of \pT, an effect specific to RHIC energies due to
the interplay between the Cronin effect and energy loss. In nuclear
collisions at the SPS (fixed target), energy loss is masked by a much
larger Cronin effect due to the much more steeply falling inclusive
spectrum, while at the LHC partonic energy loss overwhelms other
nuclear effects. The clear separation of Cronin enhancement from
energy loss at RHIC awaits the upcoming data from d+Au collisions.

\subsection{High \pT: Elliptic Flow}

Non-central nuclear collisions generate a transversely asymmetric
reaction zone.  In high multiplicity events the orientation of the
reaction plane can be determined experimentally through measurement of
the azimuthal anisotropy of the reaction
products\cite{StarEllipticFlow}. The occurrence of an anisotropy in
momentum space (``elliptic flow'') is {\it prima facie} evidence of
multiple interactions amongst the reaction products, which convert
spatial into momentum-space anisotropy. Furthermore, the asymmetry is
generated early in the collision and expansion of the system will only
dilute it\cite{FlowTheory}. The agreement in detail of hydrodynamic calculations with
elliptic flow measurements\cite{StarEllipticFlow} for $\pT<2$ GeV/c
(including mass dependence, a sensitive test) argues strongly in favor
of local equilibration early in the collision.

Finite azimuthal correlations of high \pT\ hadrons with the reaction
plane can result from partonic interactions in the
medium\cite{GVWflow}. {\it A priori} the di-jet axis has no
correlation with the reaction plane orientation. If the outgoing
parton or its hadronic fragments interacts with the surrounding
medium, the energy loss will depend upon the length of matter
transversed. The consequence will be an azimuthally varying hadron
suppression, correlated with the reaction plane.

\begin{figure}
\centering
\mbox{
\subfigure{\includegraphics[width=0.48\textwidth]{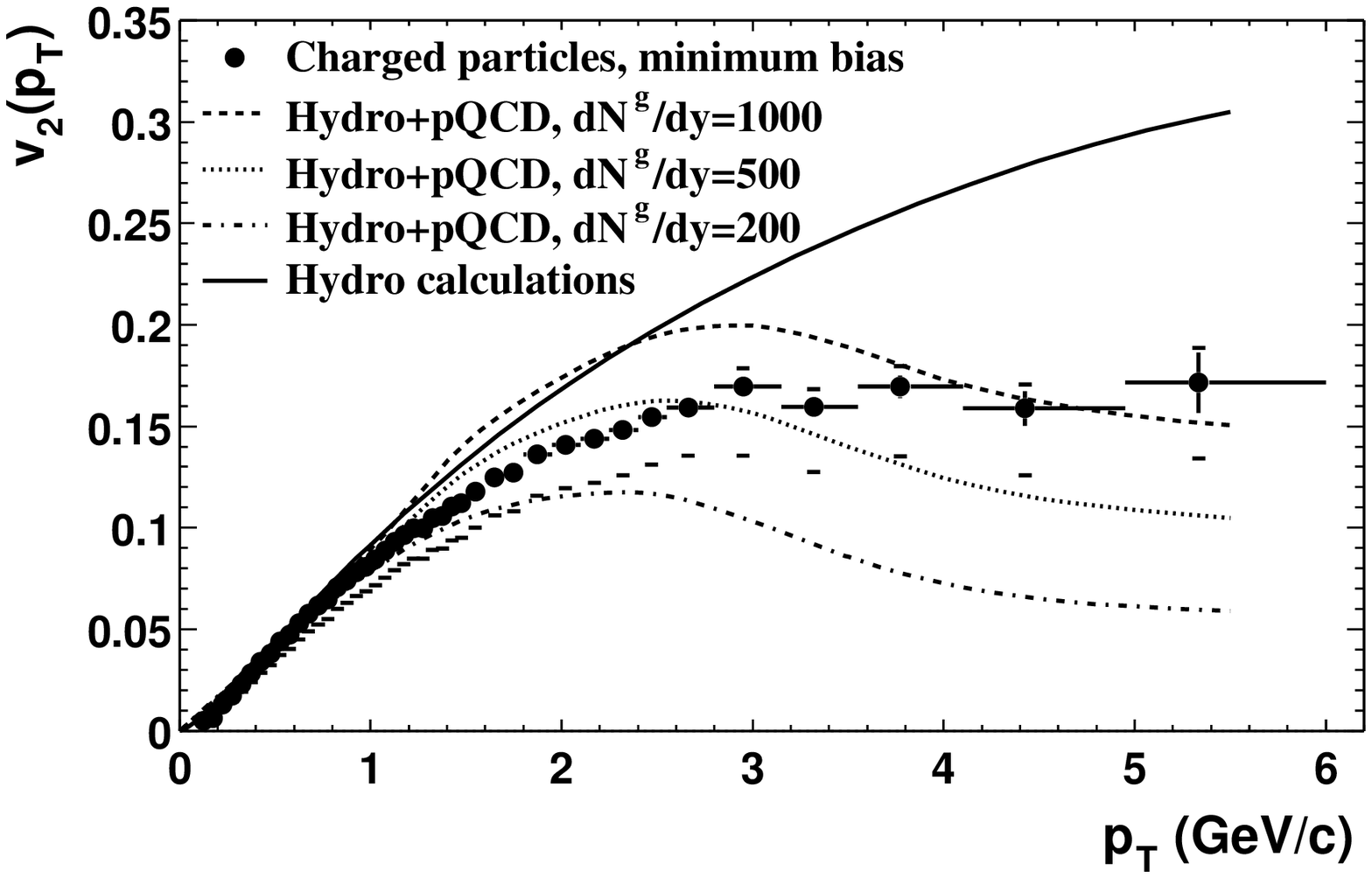}}
\subfigure{\includegraphics[width=0.48\textwidth]{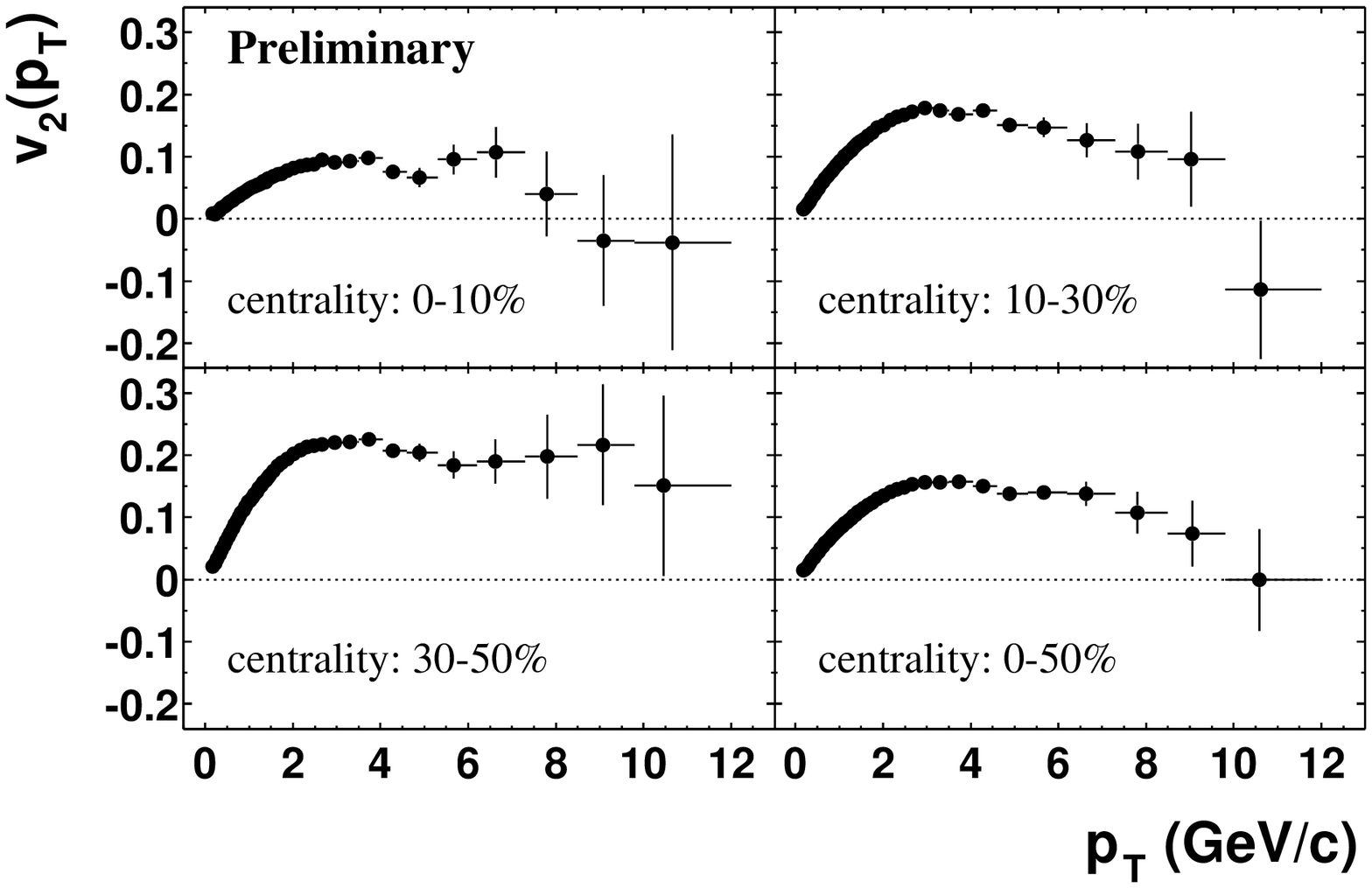}}
}
\caption{Elliptic flow \vtwo\ for charged hadrons. 
Left panel: minimum bias at 130 GeV compared to pQCD calculation\cite{STARHighptFlow,GVWflow}. 
Right panel: centrality dependence at 200 GeV\cite{KirillQM02}.}
\label{STARv2}
\end{figure}

The strength of this correlation is measured by the elliptic flow
coefficient\cite{flowapology} $\vtwo=\langle\cos(2\phi)\rangle$, the
second coefficient of the Fourier decomposition of the azimuthal
distribution relative to the reaction plane. Fig. \ref{STARv2}, left panel, shows
\vtwo\ for charged hadrons measured by STAR from minimum bias Au+Au collisions at 130
GeV, together with a calculation patching together hydrodynamics at
low
\pT\ and pQCD with partonic energy loss at high \pT\cite{GVWflow}. The 
hydrodynamic calculation agrees at $\pT<\sim2$ GeV/c. There is rough
agreement with the pQCD calculation incorporating large initial gluon
density at $\pT\sim4-6$ GeV/c. The right panel shows
\vtwo\ for the 200 GeV data, extending the measurement to
$\pT\sim12$ GeV/c. The finite correlation strength persists for all
but the most central collisions to $\pT\sim10$ GeV/c, well into the
perturbative regime. Shuryak has pointed out\cite{Shuryak} that the
the plateau values of $v_2$ for $\pT>2$ GeV/c in Fig. \ref{STARv2} are
in fact very large, exhausting (or even exceeding) the asymmetries
expected from the initial geometry of the collision.

The reaction plane for the data in Fig. \ref{STARv2} is calculated
using hadrons with $\pT<2$ GeV/c. This measurement may be biased by
multiparticle correlations unconnected to the reaction plane
(``non-flow'' effects), such as resonance decays, momentum
conservation or jet fragmentation. At $\pT\sim4$ GeV/c, non-flow
contributes $\sim$20\% to \vtwo\cite{KirillQM02}. The magnitude of
non-flow at higher \pT\ remains an open problem, though preliminary
studies indicate that the non-flow relative contribution does not
increase markedly at higher \pT.

\subsection{High \pT: Two-particle Correlations}

The two previous sections discussed inclusive hadron production and
the correlation of single leading hadrons with the reaction plane
determined from the bulk of the event ($\pT<2$ GeV/c). While jet
fragmentation is reasonably expected to dominate hadron production at
sufficiently high \pT, it remains an open question whether the current
measurements have achieved that limit. Full jet reconstruction in
Au+Au events is not possible due to the complex underlying event, but
intra-jet correlations amongst high \pT\ hadrons are still visible and
we will use these to map the transition in \pT\ from soft to hard
physics. The rate of back-to-back di-jets (di-hadrons) may be
especially sensitive to partonic energy loss effects: if one escapes
from the surface, its partner has enhanced probability to plow through
the bulk matter.

\begin{figure}
\centering
\mbox{
\subfigure[STAR\cite{STARHighptFlow,STARCorr}, large $|\Delta\eta|$ subtracted.]
{\includegraphics[width=0.58\textwidth]{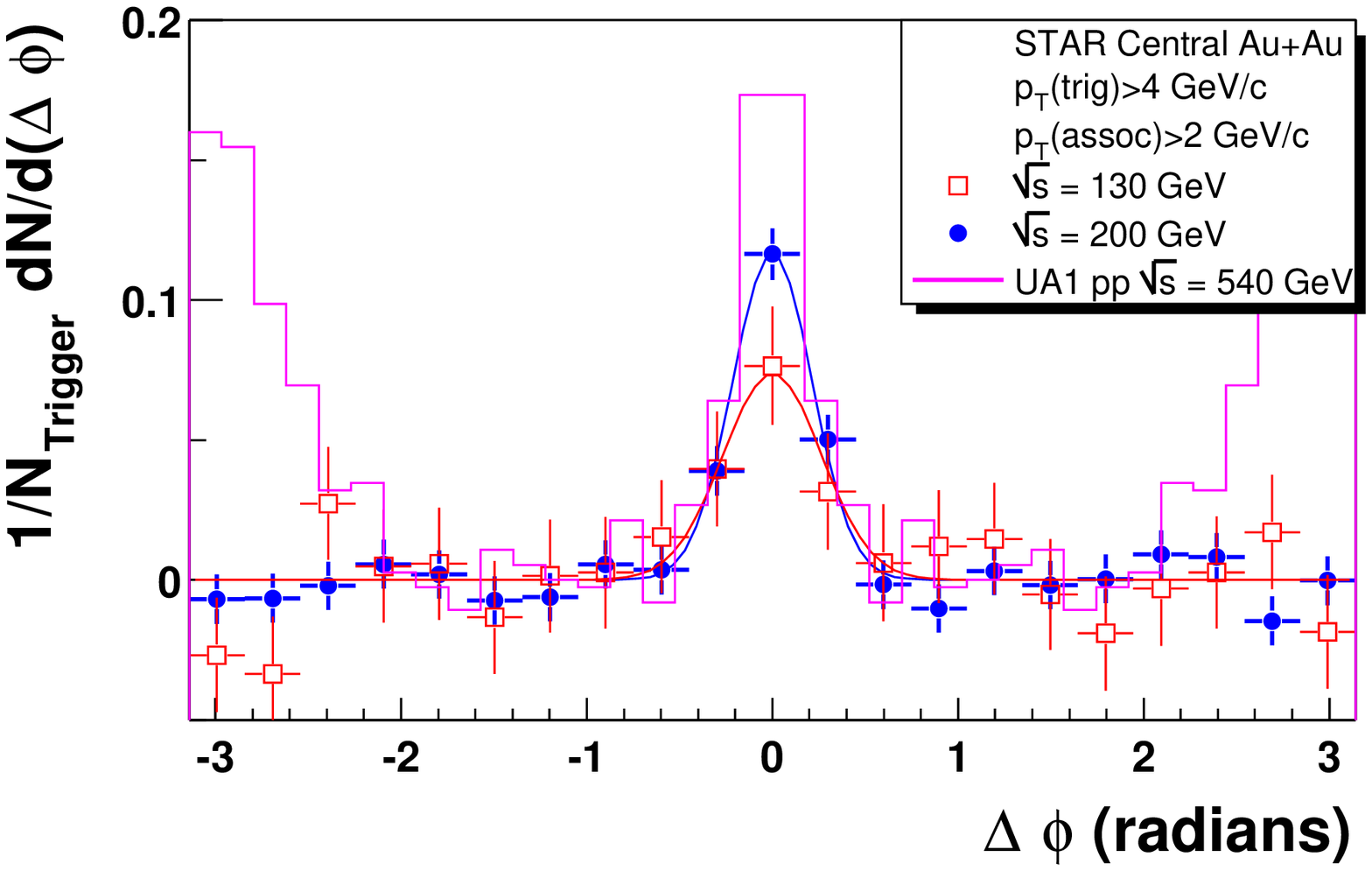}
\label{StarNearSideCorr}}
\subfigure[PHENIX\cite{SaskiaQM02}.]{\includegraphics[width=0.38\textwidth]{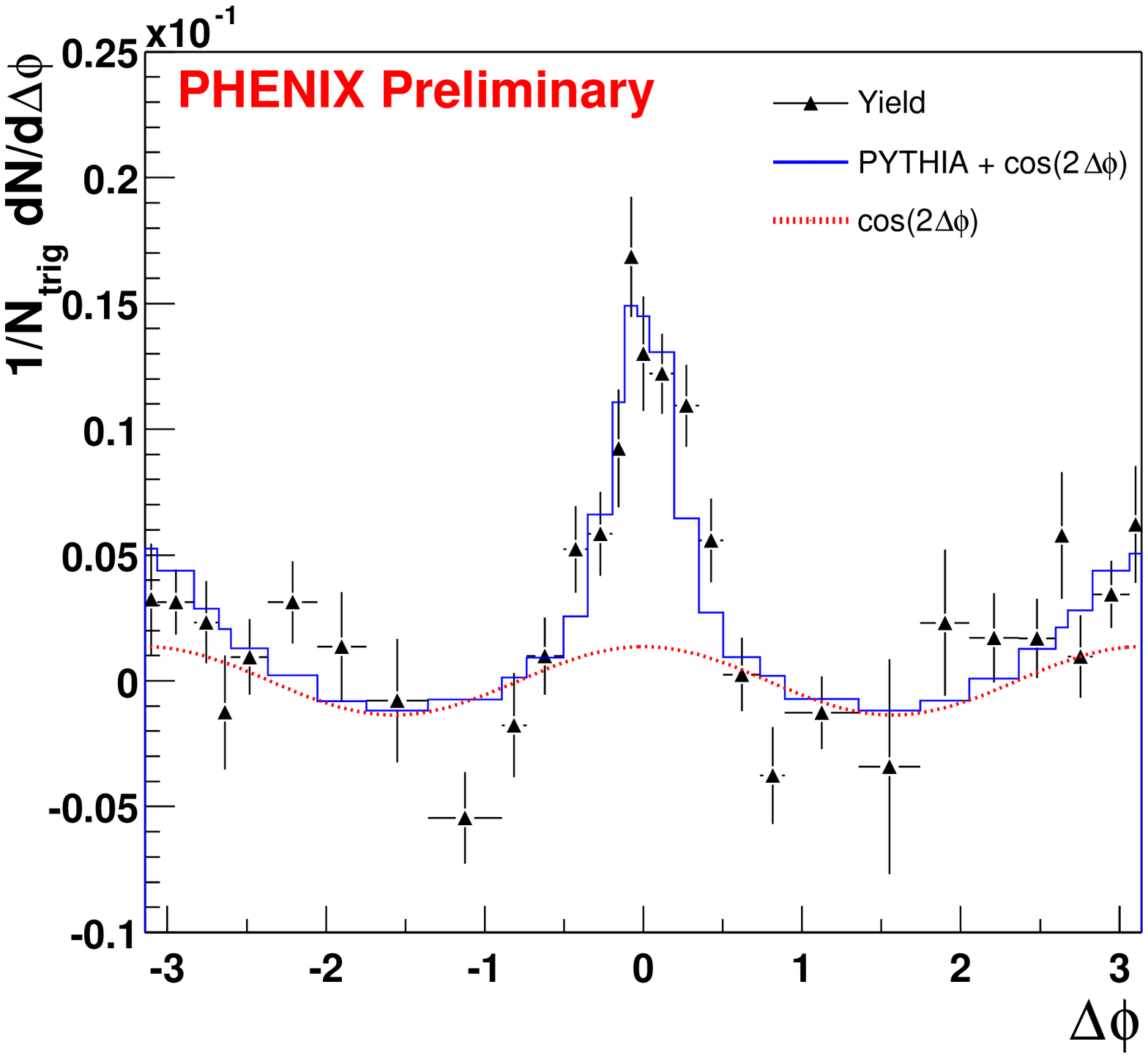}
\label{PhenixNearSideCorr}}
}
\caption{Azimuthal distributions of high \pT\ hadron pairs demonstrating intra-jet correlations. 
See text for details.}
\label{StarPhenixNearSide}
\end{figure}

\subsubsection{Near-Angle Correlations}

We first consider near-angle correlations at high \pT\ to look for
jet-like correlations. In the previous section, however, we showed that
high \pT\ hadrons are azimuthally correlated with the reaction plane
(elliptic flow) with strength $v_2$, so that even in the absence of jets,
high \pT\ hadron pairs will exhibit a correlation with strength
$v_2^2$. Elliptic flow and other non-jet contributions must be
accounted for to extract quantitatively the correlations due to
jets. 

Fig. \ref{StarPhenixNearSide} shows the distribution of relative azimuthal angle of high
\pT\ charged hadrons with respect to a trigger:

\begin{itemize}

\item Fig. \ref{StarNearSideCorr}: the trigger is a charged hadron with $\pT>4$ GeV/c
in the STAR TPC. Correlations due to non-jet effects, in particular
elliptic flow, have long range in $\Delta\eta$ (difference in
pseudorapidity between the pair), whereas intra-jet correlations have
short range in $\Delta\eta$ (caveat: also true of correlations due to
resonances). The contribution of non-jet effects to the angular
correlation can be assessed by comparing the azimuthal distributions
for pairs having $|\Delta\eta|<0.5$ and $0.5<|\Delta\eta|<1.4$. The
figure shows the difference between azimuthal distributions for small
and large $\Delta\eta$, and the peak at $\Delta\phi=0$ in the figure
is therefore short range in both $\Delta\phi$ and $\Delta\eta$. (The
away-side peak due to di-jets is subtracted by construction.)
Distributions for Au+Au collisions at 130 and 200 GeV are shown, as
well as a similar analysis from UA1 for \pbarp\ at \sqrts=540 GeV
(though without the large $|\Delta\eta|$ contribution subtracted). The
peak has similar width in all cases and grows in strength with
increasing
\sqrts, as expected due to the hardening of the underlying jet \ET\ spectrum 
with higher \sqrts.

\item Fig. \ref{PhenixNearSideCorr}: trigger is a neutral cluster in the PHENIX Electromagnetic 
Calorimeter with energy $>2.5$ GeV, principally photons from high \pT\
\pizero\ decays. The azimuthal distribution from the PYTHIA event generator matches 
correlations from p+p events measured in the same detector, assumed to
be due to jets, and PYTHIA is therefore used for comparison to
Au+Au. The dashed line shows the expectation from elliptic flow, and
the solid line shows the sum of PYTHIA and ellitpic flow. The
near-angle correlation is clearly dominated by jet contributions.

\end{itemize}

\noindent
As an additional check, the near-angle correlation exhibits a prefered
charge ordering\cite{STARCorr}, in quantitative agreement with jet
fragmentation studies at LEP\cite{DelphiChargeOrder}. The quantitative
agreement suggests that hadron prduction at $\pT>4$ GeV/c is dominated
by jet fragmentation in both p+p and Au+Au.

\subsubsection{Back-to-Back Hadron Correlations}

\begin{figure}
\centering
\mbox{
\subfigure{\includegraphics[width=0.48\textwidth]{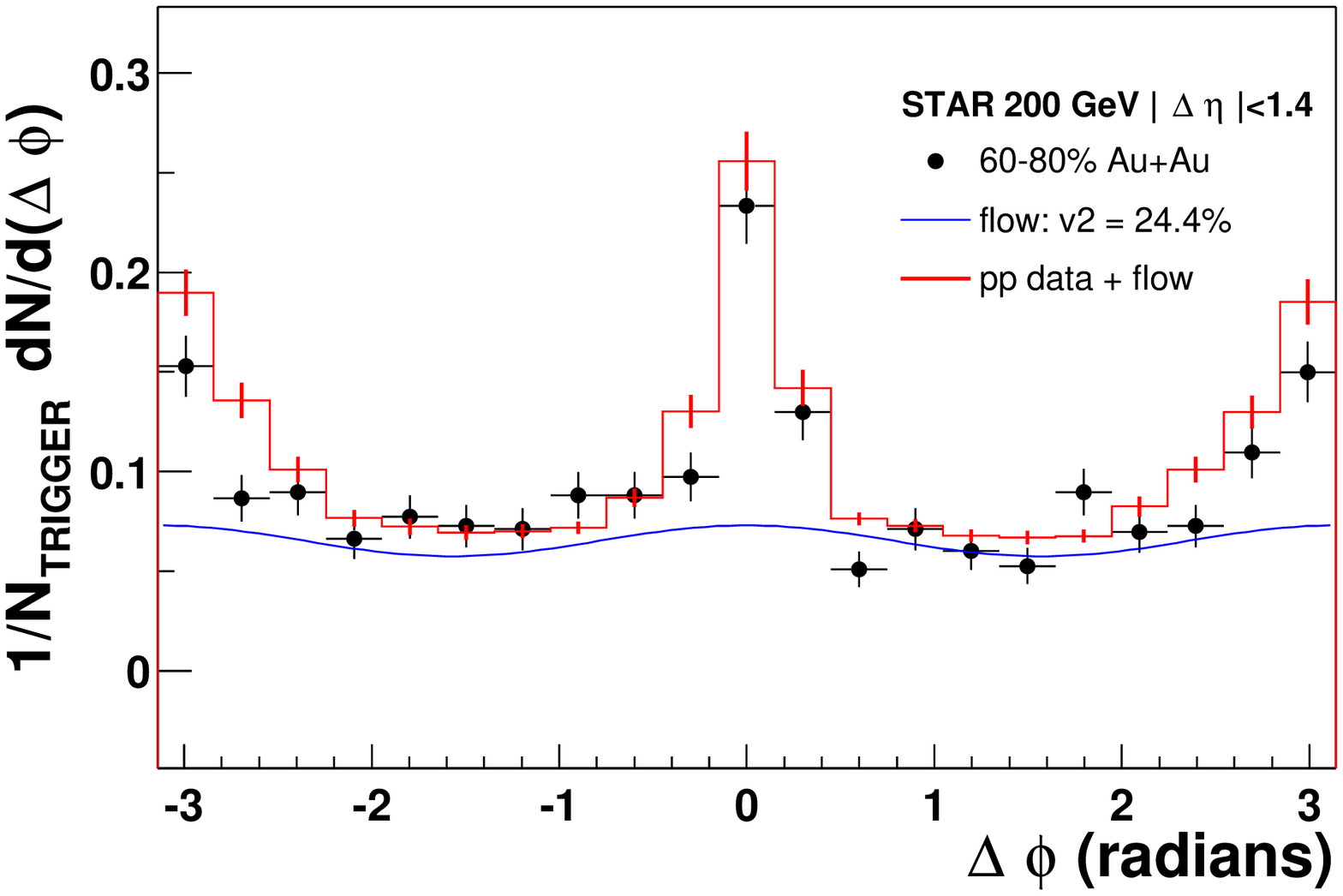}}
\subfigure{\includegraphics[width=0.48\textwidth]{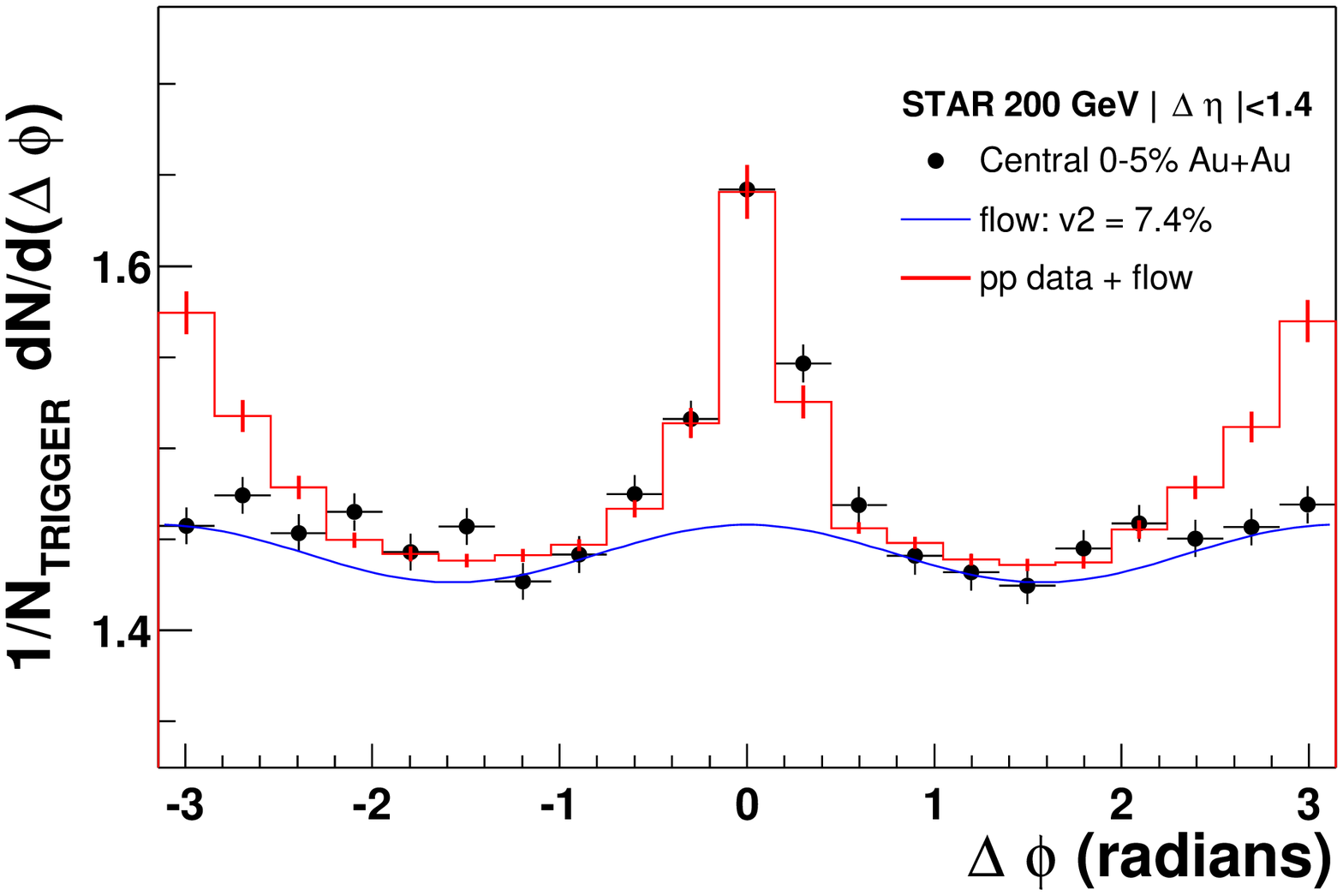}}
}
\caption{Azimuthal correlations of high \pT\ hadron pairs in Au+Au compared to p+p plus 
elliptic flow (eq. \ref{eqD2})\cite{STARCorr}. Left panel: peripheral
collisions. Right panel: central collisions.}
\label{StarAwaySide}
\end{figure}

We now turn to high \pT\ hadron pairs with large azimuthal angle
difference (back-to-back pairs) and search for the effect of partonic
energy loss on the di-jet (di-hadron) production
rate\cite{STARCorr}. Fig. \ref{StarAwaySide} shows the azimuthal
angular distribution between pairs for the most central and peripheral
Au+Au collisions at 200 GeV. The contribution of elliptic flow and
other non-jet effects must again be accounted for. Since we are
interested in the correlation strength of back-to-back pairs, which
have broad correlation in $\Delta\eta$, the trick used for the
near-angle correlations of subtracting the large $\Delta\eta$
distribution will not work. Instead, STAR compares the full azimuthal
distribution from p+p and Au+Au events measured in the same
detector. The null hypothesis is that high \pT\ correlations in Au+Au
correspond to an {\it incoherent} superposition of the jet-like
correlations measured in p+p and elliptic flow. This is expressed in
the following ansatz for the azimuthal distribution in Au+Au:

\begin{equation}
D_2^{\mathrm{AuAu}} = D_2^{\mathrm{pp}} + B(1+2v_2^2 \cos(2\Delta\phi)).
\label{eqD2}
\end{equation}

\noindent
$D_2^{\mathrm{pp}}$ is the azimuthal distribution measured in p+p,
$v_2$ is determined independently (see Fig. \ref{STARv2}) and is taken
as constant for $\pT>2$ GeV/c, and $B$ is an arbitrary normalization
to account for the combinatorial background, fit in the non-jet region
$\Delta\phi\sim\pi/2$.

\begin{wrapfigure}{l}{0.6\textwidth}
\vspace{-3mm}
\includegraphics[width=0.58\textwidth]{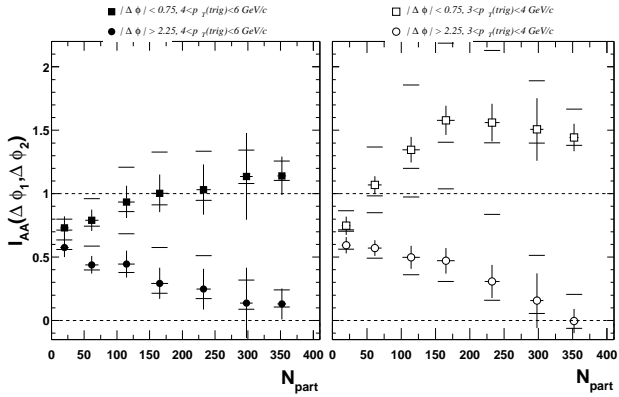}
\vspace{-3mm}
\caption{$I_{AA}$ vs centrality (large \Npart=central collisions) for near-angle (upper) and 
back-to-back (lower) angular correlations\cite{STARCorr}. Left: trigger $\pT>4$ GeV/c. 
Right: trigger $\pT>3$ GeV/c.}
\vspace{-3mm}
\label{StarIAA}
\end{wrapfigure}

The solid lines in fig. \ref{StarAwaySide} show the elliptic flow
component and the histograms are the result of eq. \ref{eqD2}. At all
centralities, the near-angle peak is fit well by eq. \ref{eqD2}, as is
the large angle peak for peripheral collisions (but slightly less
well, see below). Thus, back-to-back hadron pair production in
peripheral nuclear collisions is well described by an incoherent
superposition of elliptic flow and jet-like correlations measured in
p+p. However, the large angle peak in the most central collisions is
not described by eq. \ref{eqD2}. We are therefore led to a striking
observation: back-to-back jet production is strongly suppressed in the
most central Au+Au collisions.

The full centrality dependence of this effect is seen by subtracting
the elliptic flow component and normalizing the integrated correlation
peaks in Au+Au to those in p+p:

\begin{equation}
I_{AA}(\Delta \phi_1,\Delta \phi_2) = 
\frac{\int_{\Delta \phi_1}^{\Delta \phi_2} d(\Delta \phi) [D_2^{\mathrm{AuAu}}
- B(1+2v_2^2 \cos(2 \Delta \phi))]}{\int_{\Delta \phi_1}^{\Delta \phi_2} d
(\Delta \phi) D_2^{\mathrm{pp}}}.
\label{eqIAA}
\end{equation}

\noindent
Fig. \ref{StarIAA} shows $I_{AA}$ as a function of event
centrality for two sets of trigger and associated particle
thresholds. The near-angle peak strength is near or above unity for all
centralities and both thresholds: there is no suppression in the
near-angle correlation. In contrast, the back-to-back correlation
strength in Au+Au relative to p+p decreases smoothly from peripheral
to central collisions. For the most central collisions the strength is
consistent with zero, which would correspond to a complete suppression
of the away-side hadron correlation rate above the threshold.

Various nuclear effects may contribute to the features seen in
fig. \ref{StarIAA}. Initial state multiple scattering might generate
both the increase of the near-angle and suppression of the
back-to-back correlation strength for more central collisions. The
suppression in the most peripheral bin for Au+Au relative to p+p for
both the near-angle and back-to-back peaks could be due to nuclear
shadowing, or possibly the interplay between multiple scattering and
absorption in matter. These issues will be clarified by the upcoming
d+Au run at RHIC.

\subsection{High \pT: Baryon Enhancement}

\begin{figure}
\includegraphics[width=0.48\textwidth]{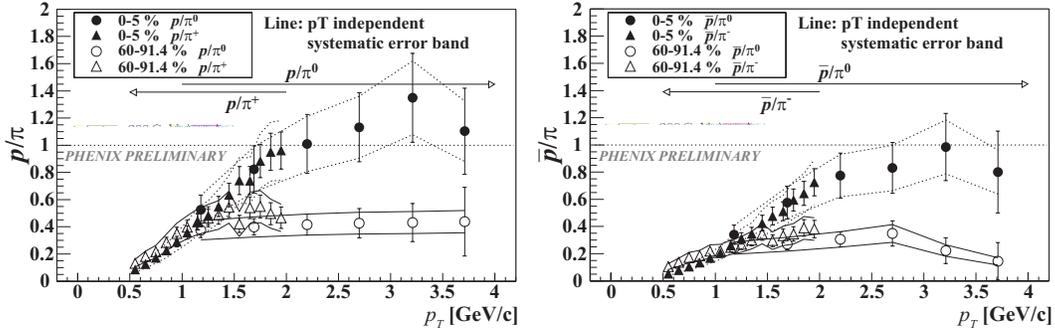}
\caption{Ratio of proton to pion yield vs \pT\ for central (filled) and 
peripheral (open) collisions\cite{SaskiaQM02}.}
\label{PhenixPOverPi}
\end{figure}

\begin{wrapfigure}{r}{0.5\textwidth}
\includegraphics[width=0.48\textwidth]{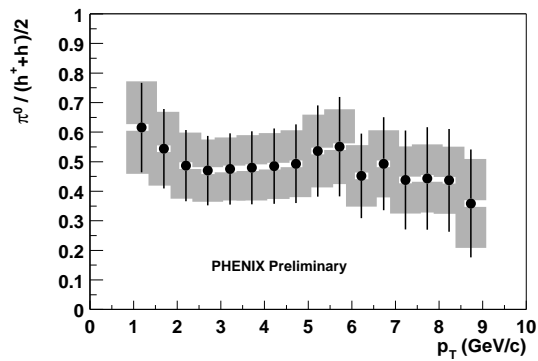}
\caption{Ratio of \pizero\ to total charged hadron yield vs. \pT\
fro minimum bias Au+Au collisions at 200 GeV\cite{SaskiaQM02}.}
\label{Phenixpi0OverCharged}
\end{wrapfigure}

Near-side azimuthal correlations indicate that hadron production at
$\pT>\sim4$ GeV/c is dominated by jet fragmentation at all
centralities. However, Fig. \ref{PhenixPOverPi} and
\ref{Phenixpi0OverCharged} from the PHENIX Collaboration showing 
the relative populations of mesons and baryons at high \pT\ present a
challenge to this picture. In fig. \ref{PhenixPOverPi} the proton/pion
ratios for peripheral collisions are in the vicinity of 0.3-0.4,
consistent with many measurements of jet fragmentation in elementary
collisions, but the ratios for central collisions are near or exceed
unity for $\pT\sim2-4$ GeV/c\cite{SaskiaQM02}, very unlike
conventional jet fragmentation in vacuum. This could be due to
transverse radial flow (hydrodynamic behavior), which has a strong
mass dependence\cite{UllrichQM02}, extending ``soft'' physics into the
nominally pertrubative regime. A second possibility is the Cronin
effect: significant enhancement of baryon relative to meson yields for
\pT$\sim$few GeV/c has been measured in fixed target p+A
measurements\cite{pABaryonMeson} and may be attributable to initial
state multiple scattering, though this measurement must be repeated at
RHIC energies for quantitative comparison. More exotic mechanisms such
as baryon production via gluonic junctions\cite{VitevJunctions} have
also been discussed in this context.

The ratios in fig. \ref{PhenixPOverPi} are measured to $\pT\sim4$
GeV/c. It is important to establish whether the anomalous baryon
enhancement for central collisions extends to higher \pT\ or whether
the conventional jet fragmentation ratio is re-established. As a step
in that direction, fig. \ref{Phenixpi0OverCharged} shows the ratio of
\pizero\ to all charged hadrons for $\pT<9$ GeV/c, for minimum bias
Au+Au events. Even for minimum bias the ratio is suppressed relative
to the conventional expectation of 0.7-0.8. This presents an important
puzzle, but more detailed centrality dependence and explicit
measurement of the baryon and meson fractions would help to 
ellucidate this phenomenon.

\subsection{High \pT: Discussion} The foregoing sections reveal the 
following facts about high \pT\ hadron production in Au+Au collisions at RHIC energies:

\begin{itemize}

\item Inclusive hadron production is suppressed in central collisions by a factor 4-5, with no strong \pT\
dependence within $5<\pT<12$ GeV/c.

\item For non-central collisions, finite azimuthal correlations of leading hadrons with the 
reaction plane persist to $\pT\sim10$ GeV/c.

\item Near-angle two particle correlations show clear jet-like correlations for $\pT>4$ GeV/c.

\item Back-to-back two particle correlations show a striking suppression in central collisions.

\item The baryon/meson ratio for $\pT<4$ GeV/c is strongly enhanced in central collisions, 
whereas it is consistent with jet fragmentation in vacuum for peripheral collisions.

\end{itemize}

\noindent
All but the last of these phenomena suggest a picture
in which the system generated in nuclear collisions at RHIC is largely
opaque to high energy partons and only those jets produced on the
periphery of the reaction zone and heading outwards survive and are
observed. This scenario naturally generates the suppression of the
inclusive spectra and the space-mometum correlation that produces
\vtwo, as well as the strong suppression of back-to-back pairs.

\begin{wrapfigure}{r}{0.65\textwidth}
\vspace{-3mm}
\includegraphics[width=0.63\textwidth]{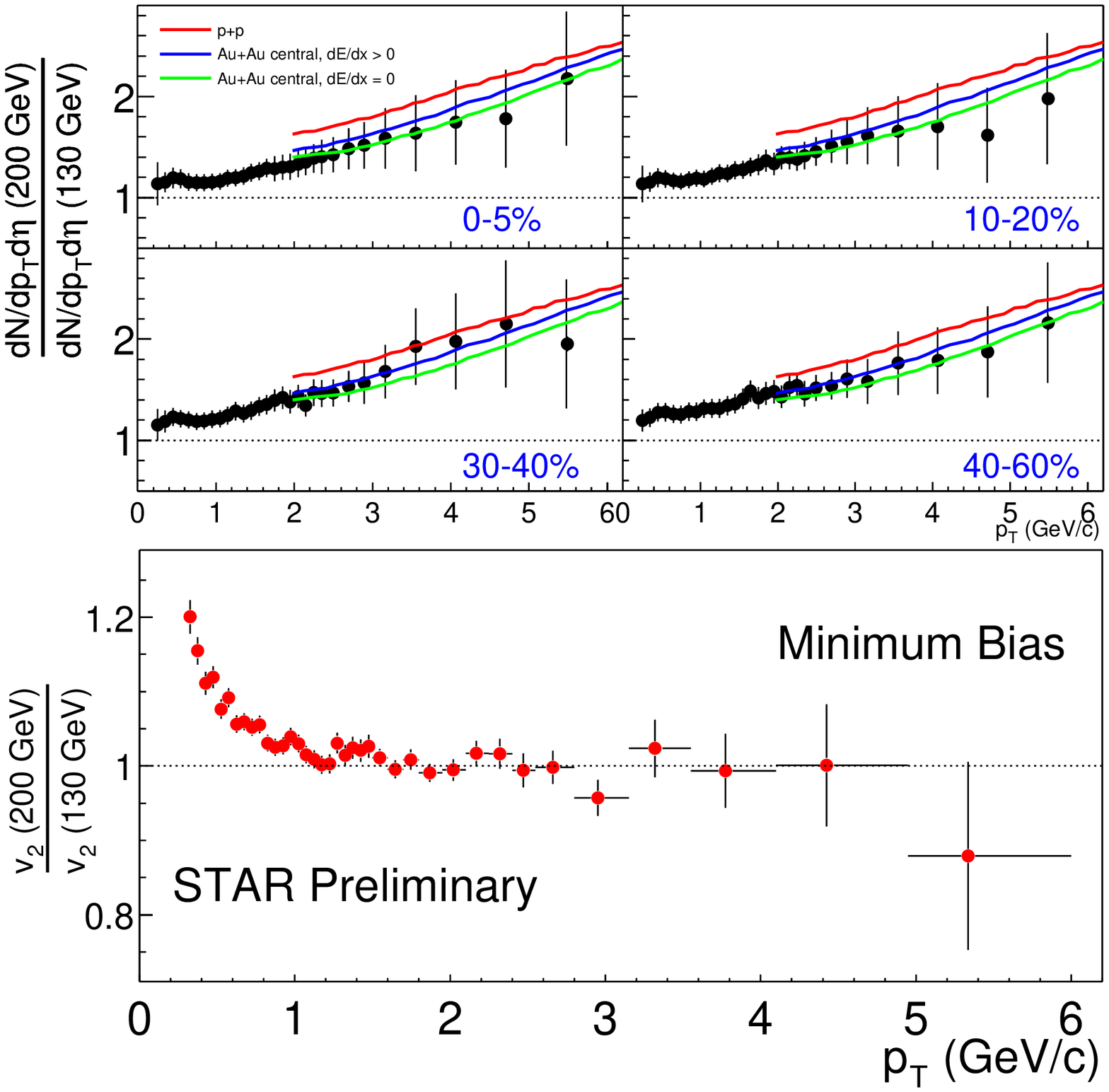}
\vspace{-3mm}
\caption{\sqrts\ dependence of inclusive yields and \vtwo\cite{JennQM02,KirillQM02}.}
\vspace{-3mm}
\label{StarSqrts}
\end{wrapfigure}
 
Additional support for this picture comes from Fig. \ref{StarSqrts},
which shows the \sqrts\ dependence of the inclusive yields and \vtwo\
between \sqrtsNN=130 and 200 GeV. The lines in the upper figure show
the pQCD expectation of the growth of inclusive cross section with
\sqrts\ for p+p and Au+Au (with and without energy
loss). The inclusive rates grow by a factor 2 at $\pT\sim6$ GeV/c,
consistent with expectations from pQCD, whereas
\vtwo\ is independent of \sqrtsNN\ at the 5\% level for $\pT>2$
GeV/c. The inclusive rates grow but the anisotropy does not change,
suggesting that the origin of \vtwo\ is geometric, not dynamic.

However, this picture as presented is only qualitative, and it remains
to be shown whether a surface emission model can simultaneously
describe the inclusive suppression, \vtwo, and the suppression of
back-to-back pairs. The anomalous baryon/meson ratio presents a puzzle
and does not seem to fit into this picture. That effect may be due to
conventional mechanisms such as transverse radial flow or initial
state multiple scattering, or more exotic physics such as baryon
production via gluonic junctions\cite{VitevJunctions}. Further
analysis of existing data and measurements of d+Au collisons at RHIC
are needed to clarify these issues.

Finally, it is necessary to establish what, precisely, is interacting
with the medium and being absorbed: partons or hadrons? Simple
estimates of hadron formation time give $t_f\sim(E/m)r$ for a hadron
of radius $r$\cite{tformation}, so that a pion with $\pT\sim5$ GeV/c
is formed well outside of the hot and dense medium. Alternatively, string
interactions in matter prior to fragmentation lead to suppression
the leading particle yields and a much shorter apparent formation
time for high \pT\ hadrons\cite{Kopeliovich}.

\section{Summary and Outlook}

RHIC has had two successful data-taking periods. It has met its design
luminosity goal for Au+Au, and has operated as the world's first
polarized proton collider. Upcoming runs in 2003 will study asymmetric
(d+Au) collisions for the first time in a collider, together with an
extended polarized proton run. A long Au+Au run at top energy will
occur in 2004, possibly with an energy scan at lower integrated
luminosity.

An extensive array of measurements in the soft physics sector indicate
the production of an equilibrated system at high temperature and pressure
which blows apart rapidly. In the high \pT\ sector, striking signals
have been observed that suggest strong partonic energy loss resulting
from a system at high energy density at early time, though there
remain important puzzles and open questions.

I have not discussed the study of heavy quark production, which is in
its infancy at RHIC. Charmonium suppression, a promising signature of
deconfinement\cite{MullerHarris}, has been observed in fixed-target
nuclear collisons at the SPS\cite{NA50}. PHENIX has measured the
inclusive electron spectrum, due primarily to
charm\cite{PHENIXDmeson}, and recently reported the first measurement
of \jpsi\ at RHIC\cite{Nagle}. This physics will be a main focus of
future runs.


\def\etal{\mbox{$\mathrm{\it et\ al.}$}} 

\newcommand{\hepph}[1]{{hep-ph/#1}}
\newcommand{\heplat}[1]{{hep-lat/#1}}
\newcommand{\hepex}[1]{{hep-ex/#1}}
\newcommand{\nuclth}[1]{{nucl-th/#1}}
\newcommand{\nuclex}[1]{{nucl-ex/#1}}
\newcommand{\astroph}[1]{{astro-ph/#1}}

\newcommand{\plb}[3]{Phys. Lett. {\bf B#1}, #3 (#2)}
\newcommand{\npa}[3]{Nucl. Phys. {\bf A#1}, #3 (#2)}
\newcommand{\npb}[3]{Nucl. Phys. {\bf B#1}, #3 (#2)}

\newcommand{\prl}[3]{Phys. Rev. Lett. {\bf #1}, #3 (#2)}
\newcommand{\prc}[3]{Phys. Rev. {\bf C#1}, #3 (#2)}
\newcommand{\prd}[3]{Phys. Rev. {\bf D#1}, #3 (#2)}

\newcommand{\nim}[3]{Nucl. Instr. Meth. {\bf #1}, #3 (#2)}

\newcommand{\arnps}[3]{Annu. Rev. Nucl. Part. Sci. {\bf B#1}, #3 (#2)}



\begin{thebibliography}{99}

\bibitem{CollinsPerry} J. C. Collins and M. J. Perry, Phys. Rev. Lett. {\bf 34}, 1353 (1975).

\bibitem{WilczekRajugopal} K. Rajagopal and F. Wilczek, \hepph{0011333}. 

\bibitem{KanayaLattice} K. Kanaya, \hepph{0209116}.

\bibitem{Karsch} F. Karsch, \npa{698}{2002}{199c}.

\bibitem{FodorQM02} Z. Fodor, \heplat{0209191}.

\bibitem{Boyanovsky} D. Boyanovsky, \hepph{0102120}.

\bibitem{GlendenningWeber} N. K. Glendenning and F. Weber, \astroph{0003426}. 

\bibitem{MullerHarris} J. W. Harris and B. M\"uller, \arnps{46}{1966}{71}.

\bibitem{Roser} T. Roser, \npa{698}{2002}{23c}.

\bibitem{RHICExperiments} BRAHMS (www4.rcf.bnl.gov/brahms/WWW/brahms.html); 
PHOBOS (phobos-srv.chm.bnl.gov); PHENIX (www.phenix.bnl.gov); STAR
(www.star.bnl.gov).

\bibitem{QM01} Proceedings of Quark Matter 2001, Stony Brook, N.Y., 
Jan. 15-21, 2001, Nucl.\ Phys.\ A698 (2002).

\bibitem{QM02} Quark Matter 2002, Nantes, France, July 11-18 2002 
(http://alice-france.in2p3.fr/qm2002/).

\bibitem{ZDC} C. Adler \etal, \nim{A461}{2001}{337}

\bibitem{PHENIXmult} K. Adcox \etal\ (PHENIX Collaboration), \prl{86}{2001}{3500}.

\bibitem{Kharzeev} D. E. Kharzeev and J. Raufeisen, \nuclth{0206073}.

\bibitem{PhobosdNchdeta} B. B. Back \etal\ (PHOBOS Collaboration), \nuclex{0210015}.

\bibitem{eta} Studies of inclusive reactions use the kinematic variables
transverse momentum \pT, rapidity $y=\frac{1}{2}\log(\frac{E+p_z}{E-p_z})$, 
and its relativistic limit pseudorapidity $\eta=-\log(\tan(\theta/2))$.

\bibitem{UllrichQM02} T. Ullrich, \nuclex{0211004}.

\bibitem{BjorkenHydro} J. D. Bjorken, \prd{27}{1983}{140}.

\bibitem{PhenixET} K. Adcox \etal\ (PHENIX Collaboration), \prl{87}{2001}{052301} 

\bibitem{PhobosSaturation} B. Back \etal\ (PHOBOS Collaboration), \prc{65}{2002}{061901R}.

\bibitem{SteinbergQM02} P. Steinberg \etal\ (PHOBOS Collaboration), \nuclex{0211002}.

\bibitem{PBMChemical} P. Braun-Munzinger, D. Magestro, K. Redlich and J. Stachel, \plb{518}{2001}{41}.

\bibitem{BjFermilabNote} J. D. Bjorken, FERMILAB-Pub-82/59-THY.

\bibitem{JetQuench} M. Gyulassy and M. Pl\"{u}mer, \plb{243}{1990}{432}; 
R. Baier \etal, \plb{345}{1995}{277}.

\bibitem{Baier} R. Baier, D. Schiff and B. G. Zakharov, \arnps{50}{2000}{37}.

\bibitem{XNfragmentation} X. N. Wang and M. Gyulassy, \prl{68}{1992}{1480}; 
X. N. Wang, \prc{58}{1998}{2321}.

\bibitem{ColdMatter} E. Wang and X. N. Wang, \prl{89}{2002}{162301}; F. Arleo, \plb{532}{2002}{231}.

\bibitem{highpT130} K. Adcox \etal\ (PHENIX Collaboration), \prl{88}{2002}{022301}; 
C. Adler \etal\ (STAR Collaboration), \prl{89}{2002}{202301}.

\bibitem{SaskiaQM02} S. Mioduszewski \etal\ (PHENIX Collaboration), \nuclex{0210021}.

\bibitem{JennQM02} J. Klay \etal\ (STAR Collaboration), \nuclex{0210026}.

\bibitem{VitevRAA} I. Vitev and M. Gyulassy, \hepph{0209161}.

\bibitem{FlowTheory} H. Sorge, \prl{82}{1999}{2048}; 
P. Kolb, J. Sollfrank and U. Heinz, \prc{62}{2000}{054909}.

\bibitem{StarEllipticFlow} C. Adler \etal\ (STAR Collaboration), \prl{86}{2001}{402}; 
\prl {87}{2001}{182301}.

\bibitem{GVWflow} M. Gyulassy, I. Vitev and X. N. Wang, \prl{86}{2001}{2537}.

\bibitem{flowapology} Use of the term ``elliptic flow'' to describe the 
azimuthal correlation of jet products with the reaction plane
orientation is unfortunate but firmly entrenched in the literature.

\bibitem{Shuryak} E. V. Shuryak, \nuclth{0112042}.

\bibitem{STARHighptFlow} C. Adler \etal\ (STAR Collaboration), \nuclex{0206006}.

\bibitem{KirillQM02} K. Filimonov \etal\ (STAR Collaboration), \nuclex{0210027}.

\bibitem{STARCorr} C. Adler \etal\ (STAR Collaboration), \nuclex{0210033}.

\bibitem{DelphiChargeOrder} P. Abreu \etal\ (DELPHI Collaboration), \plb{407}{1997}{174}.

\bibitem{pABaryonMeson} P. B. Straub \etal, \prl{452}{1992}{68}.

\bibitem{VitevJunctions} I. Vitev and M. Gyulassy, \prc{65}{2002}{041902}.

\bibitem{tformation} Y. L. Dokshitzer, V. A. Khoze, A. H. Mueller and S. I Troyan, 
``Basics of Perturbative QCD'', Editions Fronti\`eres, Gif-Sur-Yvette, 1991.

\bibitem{Kopeliovich} B. Z. Kopeliovich, \plb{243}{1990}{141}.

\bibitem{NA50} M. C. Abreu \etal\ (NA50 Collabortaion), \npa{698}{2002}{127c}.

\bibitem{PHENIXDmeson} K. Adcox \etal\ (PHENIX Collaboration), \prl{88}{2002}{192302}.

\bibitem{Nagle} J. Nagle \etal\ (PHENIX Collaboration), \nuclex{0209015}.

\end{thebibliography}
\end{document}